\documentclass{book}
\usepackage{epsfig}
\usepackage{amsmath}
\usepackage{amsfonts}
\usepackage{amssymb}
\usepackage{euscript}
\usepackage{graphicx}
\usepackage{wrapfig}
\usepackage{multicol}
\usepackage{subeqnar}
\usepackage{psfrag}
\usepackage{lscape}
\usepackage{pstricks}
\usepackage{multirow}

\begin{document}
\sloppy

\begin{center}
{{\Huge \bfseries The comparison of theoretical predictions with
measuring data of stellar parameters}}
\end{center}
\vspace{4cm}
\begin{center}
{\Huge
\itshape{ B.V.Vasiliev}}
\end{center}

\vspace{4cm}
\begin{center}
{\Huge {2006}}
\end{center}

\tableofcontents \clearpage

\chapter{Introducion} \label{Ch1A}

\section{That is a base of astrophysics?}

The empirical testing of a scientific hypothesis is the main
instrument of the present exact science. It was founded by G.Galileo
and his contemporaries. False scientific statements weren't afraid
of an empirical testing up to that time. A flight of fancy was far
refined  than an ordinary and crude material world. The exact
correspondence of philosophical theory to a check experiment was not
necessary, it almost discredits the theory in an esoteric opinion.
The discrepancy of a theory and observations was not confusing at
that time. Now the empirical testing of all theoretical hypotheses
gets a generally accepted obligatory method of exact science. As a
result all basic statements of physics are sure established. The
situation in astrophysics is quite different \footnote{The modern
astrophysics has a whole series of different branches. It has to be
emphasized that almost all of they except the physics of hot stars
are exceed the bounds of this consideration; we shall use the term
"astrophysics" here and below in its initial meaning - as the
physics of stars or more tightly as the physics of a hot stars.}.

Sometimes one considers the astrophysics as a division of science
where quantities 1, 10 and 100 are equivalent, and a comparison of
astrophysical models to measured astronomical data is impossible
according to  this reason and is not required. From its initial
stage the astrophysics was developed under assumption that its
statements are practically impossible to check. A long time this
science uses surely established "terrestrial" physical laws to
obtain its results, and it was a restriction of its empirical
testing. On the base of these experimentally tested "terresrial"
laws, astrophysicists obtain their conclusions about properties and
internal structures of distant  and mysterious stars without any
hope to check these theoretical constructions. This point of view
was clearly expressed by O.Kont (France,1798-1857):

\begin{verse}
\it{ "We don't know anything  about stars, except they are exist.
Even their temperatures stay indefinite forever"

({O.Kont "The philosophical treatise about a popular astronomy",
1844.}})
\end{verse}

Later temperatures of stars and many other of their parameters were
measured, but it did not make progress in checking of astrophysical
models. It is supposed that it is enough to use surely established
physical laws to obtain correct description of a star interior. The
astrophysical community is in the firm belief that the basis of this
science is surely formulated. \footnote{This belief gives to
astrophysicists a possibility to answer,  being asked about a
foundation of their science, that it is not a measuring data, but it
is a totality of astrophysical knowledge, models of stars and their
evolution, which gives a sure in an objective character of their
adequacy (From memory, according to words of one of known European
astrophysicist).}.

But the correctness of a fundamental scientific problem can not be
based on a belief or on an intuition and must not be  solved by
voting of professionals. Fortunately, the situation was greatly
changed in the last decades of 20th century. In according to the
progress a measuring technics, astronomers obtain the important
data. They give a possibility to check a basis of astrophysical
models. This testing will be made below.

\section{The basic postulate of astrophysics}

The basic postulate of astrophysics - the Euler equation - was
formulated in a mathematical form by L.Euler in a middle of 18th
century for the  "terrestrial" effects description. This equation
determines the equilibrium condition of  liquids or gases in a
gravitational field:
\begin{equation}
\gamma {\mathbf g} = -{\mathbf \nabla} P.\label{Eu}
\end{equation}
According to it the action of a gravity forth  $\gamma {\mathbf g}$
($\gamma$ is density of substance,  ${\mathbf g}$ is the  gravity
acceleration) in equilibrium is balanced by a forth which is induced
by the pressure gradient in the substance. This postulate is surely
established and experimentally checked be "terrestrial" physics. It
is the base of an operating of  series of technical devices -
balloons, bathyscaphes and other.

All modern models of  stellar interior are obtained on the  base of
the Euler equation. These models assume that pressure inside a star
monotone increases depthward from the star surface. As a star
interior substance can be considered as an ideal gas which pressure
is proportional to its temperature and density, all astrophysical
models predict  more or less monotonous increasing of temperature
and density of the star substance in the direction of the center of
a star.

\section{The Galileo's method}
The correctness of fundamental scientific postulate has to be
confirmed by empirical testing. The basic postulate of astrophysics
- the Euler equation (Eq.(\ref{Eu})) - was formulated long before
the  plasma theory has appeared, even long before the discovering of
electrons and nuclei. Now we are sure that the star interior must
consist of electron-nuclear plasma. The Euler equation describes
behavior of liquids or gases - substances, consisting of neutral
particles. One can ask oneself: must the electric interaction
between plasmas particles  be taken into account at a reformulation
of the Euler equation, or does it give a small correction only,
which can be neglected in most cases, and how is it  accepted by
modern stellar theories?

To solve the problem of the correct choice  of the postulate, one
has  the Galileo's method. It consists of 3 steps:

{\it (1) to postulate a hypothesis about the nature of the
phenomenon, which is free from  logical contradictions;

(2) on the base of this postulate, using the standard
mathematical procedures, to conclude laws of the phenomenon;

(3) by means of empirical method to ensure, that the nature obeys
these laws (or not) in reality, and to confirm (or not) the basic
hypothesis.}
\bigskip

The use of this method gives a possibility to reject false
postulates and theories, provided there exist a necessary
observation data, of course.

The modern theory of the star interior is based on the Euler
equation Eq.(\ref{Eu}). Another possible approach is to take into
account all interactions acting in the considered system. Following
this approach one  takes into account in the equilibrium equation
the electric interactions too, because plasma particles have
electric charges. An energetic reason speaks that plasma in gravity
field must stay electroneutral and surplus free charges must be
absent inside it.The gravity action on a neutral plasma may induce a
displacements of its particles, i.e. its electric polarization,
because electric dipole moments are lowest moments of the analysis
of multipoles. If we take into account the electric polarization
plasma $\mathfrak{P}$, the equilibrium equation obtains the form:
\begin{equation}
\gamma \mathbf {g}+\frac{4\pi}{3}\mathfrak{P}\nabla\mathfrak{P}+
\mathbf {\nabla} P =0,\label{Eu-1}
\end{equation}
On the basis of this postulate and using standard procedures, the
star interior structure can be considered (it is the main substance
of this paper).

Below it will be shown that the electric force in a very hot dense
plasma can fully compensate the gravity force and the equilibrium
equation then obtains the form:
\begin{equation}
\gamma
\mathbf{g}+\frac{4\pi}{3}\mathfrak{P}\nabla\mathfrak{P}=0\label{Eu-2}
\end{equation}
and
\begin{equation}
\mathbf {\nabla} P=0\label{P-0},
\end{equation}
which points to  a radical distinction of this approach in
comparison to the standard scheme based on Eq.(\ref{Eu}).

\section[What does the  measurement data express?]{What does the the astronomic measurement data express?}
Are there actually  astronomic measurement data, which can give
possibility to  distinguish "correct" and "incorrect" postulates?
What must one do, if the direct measurement of the star interior
construction is impossible?

Are there astronomical measurement data which can clarity this
problem? Yes. They are:

1. There is the only method which gives information about the
distribution of the matter inside stars. It is the measurement of
the  periastron rotation of closed double stars. The results of
these measurements give a qualitative confirmation that the  matter
of a star is more dense near to its center.

2. There is the measured distribution of star masses. Usually it
does not have any interpretation.

3. The astronomers detect the dependencies of radii of stars and
their surface temperatures from their masses.  Clear quantitative
explanations of these dependencies are unknown.

4. The existence of these dependencies leads to a mass-luminosity
dependence. It is indicative that the astrophysical community does
not find any  quantitative explanation to  this dependence almost
100 years after as it was discovered.

5. The discovering of the solar surface oscillations   poses a new
problem that needs a quantitative explanation.

6. There are measured data showing that a some stars are possessing
by the  magnetic fields. This date can be analyzed at taking into
account of the electric polarization inside these stars.

It seems that all measuring  data are listed which can determine the
correct choice of the starting postulate. \footnote{ From this point
of view one can consider also the measurement of solar neutrino flux
as one of similar phenomenon. But its result can be interpreted
ambiguously because there are a bad studied mechanism of their
mutual conversation and it seems prematurely to use this measurement
for a stellar models checking.}

It is important to underline that all above-enumerated dependencies
are known but they don't have a quantitative  (often even
qualitative) explanation in frame of the standard theory based on
Eq.(\ref{Eu}). Usually one does not consider the existence of these
data as a possibility to check the theoretical constructions. Often
they are ignored.

It will be shown below that all these dependencies obtain a
quantitative explanation in the theory based on the postulate
Eq.(\ref{Eu-2}), which takes in to account the  electric interaction
between particles in the stellar plasma. At that all basic measuring
parameters of stars - masses, radii, temperatures -  can be
described by definite rations of world constants, and it gives a
good agreement with measurement data.

\chapter
{A hot dense plasma} \label{Ch2}

\section{The properties of a hot dense plasma}

\subsection{A hot plasma and Boltzman distribution}

Free electrons being fermions obey the Fermi-Dirac statistic at low
temperatures. At high temperatures, quantum distinctions in behavior
of electron gas disappear and it is possible to consider electron
gas as the ideal gas which obeys the Boltzmann statistics. At high
temperatures and high densities, all substances transform into
electron-nuclear plasma. There are two tendencies in this case. At
temperature much higher than the Fermi temperature
$T_F=\frac{\mathcal{E}_F}{k}$ (where $\mathcal{E}_F$ is Fermi
energy), the role of quantum effects is small. But their role grows
with increasing of the pressure and density of an electron gas. When
quantum distinctions are small, it is possible to describe the
plasma electron gas as a the ideal one. The criterium of Boltzman's
statistics applicability
\begin{equation}
T\gg\frac{\mathcal{E}_F}{k}.\label{a-6}
\end{equation}
hold true for a nonrelativistic electron gas with density $10^{25}$
particles in $cm^{3}$ at $T\gg10^6 K$.

At this temperatures, a plasma has energy
\begin{equation}
\mathcal{E}=\frac{3}{2}kTN
\end{equation}
and its EOS is the ideal gas EOS:
\begin{equation}
P=\frac{Nk T}{V}\label{a-7}
\end{equation}

At lower temperatures, it is possible to consider  electron gas as
ideal in some approximation only. The specificity of electron gas of
plasma can be taken into account if  two corrections to ideal gas
law are introduced.

The first correction takes into account the quantum character of
electrons, which obey the Pauli principle, and cannot occupy places
which are already occupied by other electrons. This correction must
be positive because it leads to an increased gas incompressibility.

Other correction takes into account the  correlation of the
screening action of charged particles inside dense plasma. It is the
so-called correlational correction. Inside a dense plasma, the
charged particles screen the fields of other charged particles. It
leads to a decreasing of the pressure of charged particles.
Accordingly, the correction for the correlation of charged particles
must be negative,because it increases the compressibility of
electron gas.

\subsection[The correction for the  Fermi-statistic]{The hot plasma energy with taking into account the correction for the  Fermi-statistic}
The energy of the electron gas  in the Boltzmann case $(kT\gg
\mathcal{E}_F)$ can be calculated using the expression of the full
energy of a non-relativistic Fermi-particle system \cite{LL}:
\begin{equation}
\mathcal{E}=\frac{2^{1/2}V m_e^{3/2}}{\pi^2 \hbar^3} \int_0^\infty
\frac{\varepsilon^{3/2}d\varepsilon} {e^{(\varepsilon-\mu_e)/kT}+1},
\label{eg}
\end{equation}
expanding it in a series. ($m_e$ is electron mass, $\varepsilon$ is
the energy of electron and $\mu_e$ is its chemical potential).

In the Boltzmann case, $\mu_e<0$ and $|\mu_e/kT|\gg 1$ and the
integrand at $e^{\mu_e/kT}\ll 1$ can be expanded into a series
according to powers $e^{\mu_e/kT-\varepsilon/kT}$. If we introduce
the notation $z=\frac{\varepsilon}{kT}$ and conserve the two first
terms of the series,  we obtain
\begin{eqnarray}
I\equiv (kT)^{5/2}\int_0^\infty\frac{z^{3/2}dz} {e^{z-\mu_e/kT}+1}
\approx \nonumber \\ \approx (kT)^{5/2} \int_0^\infty z^{3/2}
\biggl(e^{\frac{\mu_e}{kT}-z}-e^{2(\frac{\mu_e}{kT}-z)}+...
\biggr)dz
\end{eqnarray}
or
\begin{eqnarray}
\frac{I}{(kT)^{5/2}}\approx
e^{\frac{\mu_e}{kT}}\Gamma\biggl(\frac{3}{2}+1\biggr)
-\frac{1}{2^{5/2}}e^{\frac{2\mu_e}{kT}}
\Gamma\biggl(\frac{3}{2}+1\biggr)=\nonumber \\=
 \frac{3\sqrt{\pi}}{4}
e^{\mu_e/kT}\biggl(1-\frac{1}{2^{5/2}}e^{\mu_e/kT}\biggr).
\end{eqnarray}
Thus, the full energy of the hot electron gas is
\begin{equation}
\mathcal{E}=\frac{3V}{2}\frac{(kT)^{5/2}}{\sqrt{2}}\biggl(\frac{m_e}{\pi
\hbar^2}\biggr)^{3/2} \biggl(e^{\mu_e/kT}-\frac{1}{2^{5/2}}
e^{2\mu_e/kT}\biggr)
\end{equation}
Using the definition of a chemical potential (with the spin=1/2)
\cite{LL}
\begin{equation}
\mu_e= kT log \biggl[\frac{N_e}{2V}\biggl(\frac{2\pi \hbar^2}{m_e
kT}\biggr)^{3/2}\biggr]\label{chem}
\end{equation}
we obtain the full energy of the hot electron gas
\begin{equation}
\mathcal{E}_e\approx\frac{3}{2}kTN_e\left[1+\frac{\pi^{3/2}}{4}\left(\frac{a_B
e^2}{kT}\right)^{3/2} n_e\right], \label{a-16}
\end{equation}
where $a_B=\frac{\hbar^2}{m_ee^2}$ is the Bohr radius.

\subsection[The correction for correlation  of particles]{The correction  for  correlation  of charged particles in a hot plasma}

At high temperatures, the plasma particles are uniformly distributed
in space. At this limit, the energy of ion-electron interaction
tends to zero. Some correlation in space distribution of particles
arises as the positively charged particle groups around itself
preferably particles with negative charges and vice versa. It is
accepted to estimate the energy of this correlation by the method
developed by Debye-H$\ddot u$kkel for strong electrolytes \cite{LL}.
The energy of a charged particle inside plasma is equal to
$e\varphi$, where $e$ is the charge of a particle, and $\varphi$ is
the electric potential induced by other particles  on the considered
particle.

This potential inside plasma is determined by the Debye law
\cite{LL}:

\begin{equation}
\varphi(r)=\frac{e}{r} e^{-\frac{r}{r_D}}\label{vr}
\end{equation}
where the Debye radius is
\begin{equation}
r_D=\left({\frac{4\pi e^2 }{kT}~\Sigma_a n_{a}
Z^2}\right)^{-1/2}\label{a-18}
\end{equation}
For small values of ratio $\frac{r}{r_D}$, the potential can be
expanded into a series
\begin{equation}
\varphi(r)=\frac{e}{r}-\frac{e}{r_D}+...\label{rr}
\end{equation}
The following terms are converted into zero at $r\rightarrow 0$. The
first term of this series is the potential of the considered
particle. The second term
\begin{equation}
\mathcal{E}=-e^3 \sqrt{\frac{\pi}{kTV}}\left(\Sigma_a N_a
Z_a^2\right)^{3/2}
\end{equation}
is a potential induced by other particles of plasma on the charge
under consideration. And so the correlation energy of plasma
consisting of $N_e$ electrons and $(N_e/Z)$ nuclei with charge $Z$
in volume $V$ is \cite{LL}
\begin{equation}
\mathcal{E}_{corr}=-e^3 \sqrt{\frac{\pi n_e}{kT}}\left(Z+1
\right)^{3/2}N_e \label{a-20}
\end{equation}

\section[The energetically preferable state]{The energetically preferable state of a hot plasma}
\subsection[The energetically preferable density]{The energetically preferable density  of a hot plasma}
Finally, under consideration of both main corrections taking into
account the inter-particle interaction, the full energy of plasma is
given by
\begin{equation}
\mathcal{E}=\frac{3}{2}kTN_e\biggl[1+\frac{\pi^{3/2}}{4}\biggl(\frac{a_B
e^2}{kT}\biggr)^{3/2}n_e - \frac{2\pi
^{1/2}}{3}\biggl(\frac{Z+1}{kT}\biggr)^{3/2}e^3
n_e^{1/2}\biggr]\label{a-21}
\end{equation}
The plasma into a star has a feature. A star generates the energy
into its inner region and radiates it from the surface. At the
steady state of a star, its substance must  be in the equilibrium
state with a minimum of its energy. The radiation is not in
equilibrium of course and can be considered as a star environment.
The equilibrium state of a body in an environment is related to the
minimum of the function (\cite{LL}§20):
\begin{equation}
\mathcal{E}-T_o S+P_oV, \label{a-22}
\end{equation}
where  $T_o$ and $P_o$ are the temperature and the pressure of an
environment. At taking in to account that the star radiation is
going away into vacuum, the two last items can be neglected and one
can obtain the equilibrium equation of a star substance  as the
minimum of its full energy:
\begin{equation}
\frac{d\mathcal{E}_{plasma}}{dn_e}=0.\label{a-24}
\end{equation}
Now taking into account Eq.({\ref{a-21}}), one obtains that an
equilibrium condition corresponds to the equilibrium density of the
electron gas of a hot plasma
\begin{equation}
n_e^{equilibrium}\equiv{n_\star}=\frac{16}{9\pi}\frac{(Z+1)^3}{r_B^3}\approx
1.2 \cdot 10^{24} (Z+1)^3 cm^{-3},\label{eta1}
\end{equation}
It gives the electron density $\approx3\cdot 10^{25}cm^{-3}$ for the
equilibrium state of the hot plasma of helium.

\subsection[The energetically preferable temperature]{The estimation of temperature of energetically preferable state of a hot stellar plasma}
As the steady-state value of the density of a hot non-relativistic
plasma is known, we can obtain an energetically preferable
temperature of a hot non-relativistic plasma.

The virial theorem \cite{LL,VL} claims that the full energy of
particles $E$, if they form a stable system with the Coulomb law
interaction, must be equal to their kinetic energy $T$ with a
negative sign. Neglecting small corrections at a high temperature,
one can write the full energy of a hot dense plasma as
\begin{equation}
\mathcal{E}_{plasma}= U + \frac{3}{2}kTN_e = - \frac{3}{2}kT
N_e.\label{Ts11}
\end{equation}
Where $U\approx-\frac{G\mathbb{M}^2}{\mathbb{R}_0}$ is the potential
energy of the system, $G$ is the gravitational constant,
$\mathbb{M}$ and $\mathbb{R}_0$ are the mass and the radius of the
star.

As the plasma temperature is high enough, the energy of the black
radiation cannot be neglected. The full energy of the  stellar
plasma depending on the particle energy and the black radiation
energy
\begin{equation}
\mathcal{E}_{total}=-\frac{3}{2}kT N_e + \frac{\pi^2}{15}
\biggl(\frac{kT}{\hbar c}\biggr)^3 V kT
\end{equation}
at equilibrium state must be minimal, i.e.
\begin{equation}
\biggl(\frac{\partial \mathcal{E}_{total}}{\partial T}\biggr)_{N,V}
=0.\label{a-27}
\end{equation}
This condition at $\frac{N_e}{V}=n_\star$ gives a possibility to
estimate the temperature of the hot stellar plasma at the  steady
state:
\begin{equation}
\mathbb{T}_\star\approx(Z+1)\frac{\hbar c}{kr_B}\approx
10^7(Z+1)~K.\label{Ts1}
\end{equation}
The last obtained estimation can raise doubts. At  "terrestrial"
conditions, the energy of any substance reduces to a minimum at
$T\rightarrow 0$. It is caused by a positivity of a heat capacity of
all of substances. But the steady-state energy of star is negative
and its absolute value increases with increasing of temperature
(Eq.({\ref{Ts11}})). It is the main property of a star as a
thermodynamical object. This effect is a reflection of an influence
of the gravitation on a stellar substance and is characterized by a
negative effective heat capacity. The own heat capacity of a stellar
substance (without gravitation) stays positive. With the increasing
of the temperature, the role of the black radiation increases
($\mathcal{E}_{br}\sim T^4$). When its role dominates, the star
obtains a positive heat capacity. The energy minimum corresponds to
a point between these two branches.
\subsection{Are accepted assumptions correct?}
At expansion in series of the full energy of a Fermi-gas, it was
supposed that the condition of applicability of Boltzmann-statistics
({\ref{a-6}}) is valid. The substitution of obtained values of the
equilibrium density  $n_\star$ ({\ref{eta1}}) and equilibrium
temperature $\mathbb{T}_\star$ ({\ref{Ts1}}) shows that the ratio
\begin{equation}
\frac{\mathcal{E}_F(n_\star)}{k\mathbb{T}_\star}\approx
3.1(Z+1)\alpha.\label{alf}.
\end{equation}
Where $\alpha$ is fine structure constant.

The condition ($\frac{r}{r_D}\ll1$), used at expansion in series of
the electric potential  near a nucleus ({\ref{rr}}), gives at
appropriate substitution
\begin{equation}
n_\star^{1/3}r_D\approx {\alpha}^{-1/2}.
\end{equation}
Thus, obtained values of steady-state parameters of plasma are in
full agreement with assumptions which was made above.

\chapter[The gravity induced electric polarization]{The gravity induced electric polarization in a dense hot plasma}\label{Ch3}

\section{Plasma cells}

The existence of plasma at energetically favorable state with the
constant density $n_\star$ and the constant temperature
$\mathbb{T}_\star$  puts a question about equilibrium of this plasma
in a gravity field. The Euler equation in commonly accepted form
Eq.({\ref{Eu}) disclaims a possibility to reach the equilibrium in a
gravity field at a constant pressure in the substance: the gravity
inevitably must induce a pressure gradient into gravitating matter.
To solve this problem, it is necessary to consider  the equilibrium
of a dense plasma in an gravity field in detail. At zero
approximation, at a very high temperature, plasma can be considered
as a "jelly", where electrons and nuclei are "smeared" over a
volume. At a lower temperature and a high density, when an
interpartical interaction cannot be neglected, it is accepted to
consider a plasma dividing in cells \cite{Le}. Nuclei are placed at
centers of these cells, the rest of their volume is filled by
electron gas. Its density decreases from the center of a cell to its
periphery. Of course, this dividing is not freezed. Under action of
heat processes, nuclei move. But having a small mass, electrons have
time to trace this moving and to form a permanent electron cloud
around nucleus, i.e. to form a cell. So plasma under action of a
gravity must be characterized by two equilibrium conditions:

- the condition of an equilibrium of the heavy nucleus inside a
plasma cell;

- the condition of an equilibrium of the electron gas, or
equilibrium of cells.

\section[The equilibrium of a nucleus]{The equilibrium of a nucleus inside plasma
cell filled by an electron gas}

At the absence of gravity, the negative charge of an electron cloud
inside a cell exactly balances the positive charge of the nucleus at
its center. Each cell is fully electroneutral. There is no  direct
interaction between nuclei.

The gravity acts on electrons and nuclei at the same time. Since the
mass of nuclei is large, the gravity force applied to them is  much
larger than the force applied to electrons. On the another hand, as
nuclei have no direct interaction, the elastic properties of plasma
are depending on the electron gas reaction. Thus er have a
situation, when the force applied to nuclei  must be balanced by the
force of the electron subsystem. The cell obtains an electric dipole
moment $d_s$, and  the plasma obtains polarization
$\mathfrak{P}=n_s~d_s$, where $n_s$ is the density of the cell.

It is known \cite{LL8}, that the polarization of neighboring cells
induces in the considered cell the electric field intensity
\begin{equation}
E_s=\frac{4\pi}{3}\mathfrak{P},
\end{equation}
and the cell obtains the energy
\begin{equation}
\mathcal{E}_s=\frac{d_s~E_s}{2}.
\end{equation}

The gravity force applied to the nucleus is proportional to its mass
$Am_p$ (where $A$ is a mass number of the nucleus, $m_p$ is the
proton mass). The cell consists of $Z$ electrons, the gravity force
applied to the cell electron gas is proportional to $Z m_e$ (where
$m_e$ is the electron mass). The difference of these forces tends to
pull apart centers of positive and negative charges and to increase
the dipole moment. The electric field  $E_s$ resists it. The process
obtains equilibrium at the balance of the arising electric force
$\nabla \mathcal{E}_s$ and the difference of gravity forces applied
to the electron gas and the nucleus:
\begin{equation}
\mathbf{\nabla}\left(\frac{2\pi}{3}\frac{\mathfrak{P}^2}{n_s}\right)+(Am_p-Zm_e)\mathbf{g}=0\label{b2}
\end{equation}
Taking into account, that $\mathbf{g}=-\nabla \psi$, we obtain
\begin{equation}
\frac{2\pi}{3}\frac{\mathfrak{P}^2}{n_s}= (Am_p-Zm_e)\psi.\label{b3}
\end{equation}
Hence,
\begin{equation}
\mathfrak{P}^2 = \frac{3GM_r}{2\pi
r}n_e\left(\frac{A}{Z}m_p-m_e\right),\label{b5}
\end{equation}
where $\psi$ is the potential of the gravitational field,
$n_s=\frac{n_e}{Z}$ is the density of cell (nuclei), $n_e$ is the
density of the electron gas, $M_r$ is the mass of a star containing
inside a sphere with radius $r$.

\section[The equilibrium in electron gas subsystem]{The equilibrium in plasma electron gas subsystem}

Nonuniformly polarized matter can be represented by an electric
charge distribution with density  \cite{LL8}
\begin{equation}
\widetilde{\varrho}= \frac{div E_s}{4\pi}=\frac{div\mathfrak{P}}{3}.
\end{equation}
The full electric charge of cells placed inside the sphere with
radius $r$
\begin{equation}
Q_r= 4\pi \int_0^r \widetilde{\varrho}r^2 dr
\end{equation}
determinants the electric field intensity applied to a cell placed
on a distance  $r$  from center of a star
\begin{equation}
\widetilde{\mathbf{E}}=\frac{Q_r}{r^2}
\end{equation}
As a result, the action of a nonuniformly polarized environment can
be described by the force $\widetilde{\varrho}\widetilde{E}$. This
force must be taken into account in the formulating of equilibrium
equation. It leads to the following form  of the Euler equation:
\begin{equation}
\gamma\mathbf{g}+\widetilde{\varrho}\widetilde{\mathbf{E}}+\nabla
P=0\label{Eu1}
\end{equation}

\chapter{The internal structure of a star} \label{Ch4}

It was shown above that the state with the constant density is
energetically favorable for a plasma at a very high temperature. The
plasma in the central region of a star can possess by this property
. The  calculation  made below shows that  the mass of central
region of a star with the constant density  - the star core - is
equal to 1/2 of the full star mass. Its radius is approximately
equal to 1/10 of radius of a star, i.e. the  core with high density
take approximately 1/1000 part of the full volume of a star. The
other half of a stellar matter is distributed over the region placed
above the core. It has a relatively small density and it could be
called as a star atmosphere.

\section{The plasma equilibrium in the star core}
In this case,  the equilibrium condition (Eq.(\ref{b2})) for steady
density plasma is achieved at
\begin{equation}
\mathfrak{P}=\sqrt{G}\gamma_\star r,
\end{equation}
Here the mass density is $\gamma_\star\approx\frac{A}{Z}m_p
n_\star$. The polarized state of the plasma can be described by a
state with  an electric charge at the density
\begin{equation}
\widetilde{\varrho}=\frac{1}{3}div
\mathfrak{P}=\sqrt{G}\gamma_\star,\label{roe}
\end{equation}
and the electric field applied to a cell is
\begin{equation}
\widetilde{\mathbf{E}}=\frac{\mathbf{g}}{\sqrt{G}}.
\end{equation}
As a result, the electric force applied to the cell will fully
balance the gravity action
\begin{equation}
\gamma\mathbf{g}+\widetilde{\varrho}\widetilde{\mathbf{E}}=0\label{Eu2}
\end{equation}
at the zero pressure gradient
\begin{equation}
\nabla P=0\label{EuP}.
\end{equation}

\section[The main parameters of a star core]{The main parameters of a star core (in order of values)}

At known density $n_\star$ of plasma into a core  and its
equilibrium temperature $\mathbb{T}_\star$, it is possible to
estimate the mass $\mathbb{M}_\star$ of a star core  and its radius
$\mathbb{R}_\star$. In accordance with the virial
theorem\footnote{Below we shell use this theorem in its more exact
formulation.}, the kinetic energy of particles composing the steady
system  must be approximately equal to its potential energy with
opposite sign:
\begin{equation}
\frac{G\mathbb{M}_\star^2}{\mathbb{R}_\star}\approx
k\mathbb{T}_\star\mathbb{N}_\star\label{b30}.
\end{equation}
Where $\mathbb{N}_\star=\frac{4\pi}{3}\mathbb{R}_\star^3 n_\star$ is
full number of particle into the star core.

With using determinations derived above ({\ref{eta1}) and
({\ref{Ts1})  derived before, we obtain
\begin{equation}
\mathbb{M}_\star \approx \frac{\mathbb{M}_{Ch}}{(A/Z)^2}\label{Ms}
\end{equation}
where  $\mathbb{M}_{Ch}=\left(\frac{\hbar c}{G
m_p^2}\right)^{3/2}m_p$ is the Chandrasekhar mass.

The radius of the core is approximately equal
\begin{equation}
\mathbb{R}_\star\approx\biggl(\frac{\hbar c}{G m_p^2}\biggr)^{1/2}
\frac{a_B}{(Z+1){A/Z}}.\label{RN}
\end{equation}
where  $A$ and  $Z$ are the mass and the charge number of atomic
nuclei the plasma consisting of.

\section[The equilibrium inside the star atmosphere]{The equilibrium state of the plasma inside the star atmosphere}
The star core is characterized by the constant mass density, the
charge density, the temperature and the pressure. At a temperature
typical for a star core, the plasma can be considered as ideal gas,
as interactions between its particles are small in comparison with
$k\mathbb{T}_\star$. In atmosphere, near surface of a star, the
temperature is approximately by $3\div 4$ orders smaller. But the
plasma density is lower. Accordingly, interparticle interaction is
lower too and we can continue to consider this plasma as ideal gas.

In the absence of the gravitation, the equilibrium state of ideal
gas in some volume comes with the pressure equalization, i.e. with
the equalization of its temperature $T$ and its density $n$. This
equilibrium state is characterized by the equalization of the
chemical potential of the gas $\mu$ (Eq.({\ref{chem})).

\section[The radial dependence of density and temperature]{The radial dependence of density and temperature of substance inside a star atmosphere}

For the equilibrium system, where different parts have  different
temperatures, the following relation  of the chemical potential of
particles to its temperature holds (\cite {LL},§25):
\begin{equation}
\frac{\mu}{kT}=const
\end{equation}
As thermodynamic (statistical) part of chemical potential of
monoatomic ideal gas is \cite{LL},{§45}:
\begin{equation}
\mu_T= kT~ln \biggl[\frac{n}{2}\biggl(\frac{2\pi \hbar^2}{m
kT}\biggr)^{3/2}\biggr],\label{muB}
\end{equation}
we can conclude that at the equilibrium
\begin{equation}
n\sim T^{3/2}.\label{b32}
\end{equation}
In external fields the chemical potential of a gas \cite{LL}§25 is
equal to
\begin{equation}
\mu=\mu_T + \mathcal{E}^{potential}
\end{equation}
where $\mathcal{E}^{potential}$ is the potential energy of particles
in the external field. Therefore in addition to fulfillment of
condition Eq. ({\ref{b32}), in a field with Coulomb potential, the
equilibrium needs a fulfillment of the condition
\begin{equation}
-\frac{GM_r \gamma}{r kT_r}+
\frac{\mathfrak{P}_r^2}{2kT_r}=const\label{gP}
\end{equation}
(where $m$ is the particle mass, $M_r$ is the mass of a star inside
a sphere with radius $r$, $\mathfrak{P}_r$ and $T_r$ are the
polarization and the temperature on its surface. As on the core
surface, the left part of Eq.(\ref{gP}) vanishes, in the atmosphere
\begin{equation}
M_r \sim r kT_r.\label{m-tr}
\end{equation}
Supposing  that a decreasing  of temperature inside the atmosphere
is a power function with the exponent $x$, its value on a  radius
$r$ can be written as
\begin{equation}
T_r=\mathbb{T}_\star\biggl(\frac{\mathbb{R_\star}}{r}\biggr)^x\label{Tr}
\end{equation}
and in accordance with  ({\ref{b32}}), the density
\begin{equation}
n_r=n_\star\biggl(\frac{\mathbb{R}_\star}{r}\biggr)^{3x/2}.\label{nr}
\end{equation}
Setting the  powers of $r$ in the right and the left parts of the
condition ({\ref{m-tr}) equal, one can obtain $x=4$.

Thus, at using power dependencies for the description of radial
dependencies of density and temperature, we obtain
\begin{equation}
n_r={n}_\star\biggl(\frac{\mathbb{R}_\star}{r}\biggr)^6\label{an-r}
\end{equation}
and
\begin{equation}
T_r=\mathbb{T}_\star\biggl(\frac{\mathbb{R}_\star}{r}\biggr)^4.\label{tr}
\end{equation}

\section[The mass of the star]{The mass of the star atmosphere and the full mass of a star}
After integration of ({\ref{an-r}}), we can obtain the mass of the
star atmosphere
\begin{equation}
\mathbb{M}_{A} = 4\pi \int_{\mathbb{R}_\star}^{\mathbb{R}_0} (A/Z)
m_p n_\star \biggl(\frac{\mathbb{R}_\star}{r}\biggr)^6 r^2 dr=
\frac{4\pi}{3}(A/Z)m_p n_\star
\mathbb{R_\star}^3\left[1-\left(\frac{\mathbb{R}_\star}{\mathbb{R}_0}\right)^3\right]\label{ma}
\end{equation}
It is equal to its core mass (to
$\frac{\mathbb{R}_\star^3}{\mathbb{R}_0^3}\approx 10^{-3}$), where
$\mathbb{R}_0$ is radius of a star.

Thus, the full mass of a star
\begin{equation}
\mathbb{M} = \mathbb{M}_A+ \mathbb{M}_\star\approx
2\mathbb{M}_\star\label{2mstar}
\end{equation}
The mass of the Sun core can be estimated from measuring data of the
Sun surface oscillations. It will be shown below that according to
this data  the ratio of the core mass to the full mass of the Sun
(Eq.({\ref{MrS}})) is really near to $1/2$ in according with
Eq.({\ref{2mstar}}). Let us mark a difference between determinations
of mass inside the core and inside the atmosphere. Inside the core,
in a sphere with radius $r$ the mass
\begin{equation}
M(r)= \frac{4\pi}{3}\gamma_\star r^3,\label{mr-core}
\end{equation}
is settled.

Inside the atmosphere ($\mathbb{R}_\star<r<\mathbb{R}_0$), the mass
of the spherical volume of the radius  $r$ is
\begin{equation}
M(r)=
\mathbb{M_\star}\left[2-\left(\frac{\mathbb{R}_\star}{r}\right)^3\right]\label{rma}
\end{equation}

\chapter [The main parameters of star]
{The virial theorem  and main parameters of a star} \label{Ch5}

\section{The energy of a star}

The virial theorem \cite{LL,VL} is applicable to a  system of
particles if they have a finite moving into a volume  $V$. If their
interaction obeys to the Coulomb's law, their potential energy
$\mathcal{E}^{potential}$, their kinetic energy
$\mathcal{E}^{kinetic}$ and pressure  $P$ are in the ratio:
\begin{equation}
2\mathcal{E}^{kinetic}+\mathcal{E}^{potential}= 3PV.
\end{equation}
On the star surface, the pressure is absent and for the particle
system as a whole:
\begin{equation}
2\mathcal{E}^{kinetic}=-\mathcal{E}^{potential}
\end{equation}
and the full energy of plasma particles into a star
\begin{equation}
\mathcal{E}_{star}=\mathcal{E}^{kinetic}+\mathcal{E}^{potential}=-\mathcal{E}^{kinetic}.\label{ek}
\end{equation}
Let us calculate the separate items composing the full energy of a
star.
\subsection{The kinetic energy of plasma}

The kinetic energy of plasma into a core:
\begin{equation}
\mathcal{E}_\star^{kinitic}=\frac{3}{2}k\mathbb{T}_\star
\mathbb{N}_\star.
\end{equation}
The kinetic energy of atmosphere:
\begin{equation}
\mathcal{E}_a^{kinetic}=4\pi\int_{\mathbb{R}_\star}^{\mathbb{R}_0}
\frac{3}{2}k\mathbb{T}_\star
n_\star\left(\frac{\mathbb{R}_\star}{r}\right)^{10}
r^2dr\approx\frac{3}{7}\left(\frac{3}{2}k\mathbb{T}_\star
\mathbb{N}_\star\right)
\end{equation}
The total kinetic energy of plasma particles
\begin{equation}
\mathcal{E}_{plasma}=\mathcal{E}_\star^{kinetic}+\mathcal{E}_a^{kinetic}=\frac{15}{7}k\mathbb{T}_\star
\mathbb{N}_\star\label{kes}
\end{equation}

\subsection{The potential energy of star plasma}
Inside a star core, the gravity force is balanced by the force of
electric nature. Correspondingly, the energy of electric
polarization can be considered as balanced by the gravitational
energy of plasma. As a result, the potential energy of a core can be
considered as equal to zero.

In a star atmosphere, this balance is absent.

The gravitational energy of an atmosphere
\begin{equation}
\mathcal{E}_a^{G}=-\cdot 4\pi G\mathbb{M}_\star
m'n_\star\int_{\mathbb{R}_\star}^{\mathbb{R}_0}
\frac{1}{2}\left[2-\left(\frac{\mathbb{R}_\star}{r}\right)^{3}\right]\left(\frac{\mathbb{R}_\star}{r}\right)^{6}r
dr
\end{equation}
or
\begin{equation}
\mathcal{E}_a^{G}=\frac{3}{2}\left(\frac{1}{7}-\frac{1}{2}\right)\frac{G\mathbb{M}_\star^2}{\mathbb{R}_\star}
=-\frac{15}{28}\frac{G\mathbb{M}_\star^2}{\mathbb{R}_\star}
\end{equation}

The electric energy of atmosphere is
\begin{equation}
\mathcal{E}_a^{E}=-4\pi\int_{\mathbb{R}_\star}^{R_0}
\frac{1}{2}\varrho\varphi r^2 dr,
\end{equation}
where
\begin{equation}
\widetilde{\varrho}=\frac{1}{3r^2}\frac{d\mathfrak{P}r^2}{dr}
\end{equation}
and
\begin{equation}
\widetilde{\varphi}=\frac{4\pi}{3}\mathfrak{P}r.
\end{equation}
The electric energy:
\begin{equation}
\mathcal{E}_a^{E}=-
\frac{3}{28}\frac{G\mathbb{M}_\star^2}{\mathbb{R}_\star},
\end{equation}
and total potential energy of atmosphere:
\begin{equation}
\mathcal{E}_a^{potential}=\mathcal{E}_a^{G}+\mathcal{E}_a^{E}=
-\frac{9}{14}\frac{G\mathbb{M}_\star^2}{\mathbb{R}_\star}.\label{epa}
\end{equation}
The equilibrium in a star depends both on plasma energy and energy
of radiation.

\section{The temperature of a star core}
\subsection{The energy of the black radiation}
The energy of black radiation inside a star core is
\begin{equation}
\mathcal{E}_{br}^\star= \frac{\pi^2}{15}k\mathbb{T}_\star
\left(\frac{k\mathbb{T}_\star}{\hbar c}\right)^{3}\mathbb{V}_\star.
\end{equation}
The energy of black radiation inside a star atmosphere  is
\begin{equation}
\mathcal{E}_{br}^a=4\pi\int_{R_\star}^{R_0}
\frac{\pi^2}{15}k\mathbb{T}_\star
\left(\frac{k\mathbb{T}_\star}{\hbar
c}\right)^{3}\left(\frac{\mathbb{R}_\star}{r}\right)^{16}
r^2dr=\frac{3}{13}\frac{\pi^2}{15}k\mathbb{T}_\star
\left(\frac{k\mathbb{T}_\star}{\hbar c}\right)^{3}\mathbb{V}_\star.
\end{equation}
The total energy of black radiation inside a star is
\begin{equation}
\mathcal{E}_{br}^\Sigma=\mathcal{E}_{br}^\star+\mathcal{E}_{br}^a=\frac{16}{13}\frac{\pi^2}{15}kT_\star
\left(\frac{kT_\star}{\hbar
c}\right)^{3}V_\star=1.23\frac{\pi^2}{15}kT_\star
\left(\frac{kT_\star}{\hbar c}\right)^{3}V_\star
\end{equation}

\subsection{The full energy of a star}
In accordance with  ({\ref{ek}}), the full energy of a star
\begin{equation}
\mathcal{E}_{STAR}=-\mathcal{E}_{kinetic}+\mathcal{E}_{br}^\Sigma
\end{equation}
i.e.
\begin{equation}
\mathcal{E}_{STAR}=-\frac{15}{7}k\mathbb{T}_\star
\mathbb{N}_\star+\frac{16}{13}\frac{\pi^2}{15}k\mathbb{T}_\star
\left(\frac{k\mathbb{T}_\star}{\hbar c}\right)^{3}\mathbb{V}_\star.
\end{equation}
The steady state of a star is determined by a minimum of its full
energy:
\begin{equation}
\left(\frac{d\mathcal{E}_{STAR}}{d\mathbb{T}_\star}\right)_{\mathbb{N}=const,\mathbb{V}=const}=0,
\end{equation}
it corresponds to the condition:
\begin{equation}
-\frac{15}{7} \mathbb{N}_\star+\frac{64\pi^2}{13\cdot 15}
\left(\frac{k\mathbb{T}_\star}{\hbar
c}\right)^{3}\mathbb{V}_\star=0.
\end{equation}
Together with Eq.({\ref{eta1}}) it defines the equilibrium
temperature of a star core:
\begin{equation}
\mathbb{T}_\star=\left(\frac{25\cdot
13}{28\pi^{4}}\right)^{1/3}\left(\frac{\hbar
c}{ka_B}\right)(Z+1)\approx (Z+1)\cdot 2.13\cdot 10^7 K\label{tcore}
\end{equation}

\section{Main star parameters}
\subsection{The star mass}
The virial theorem relates kinetic energy of a system with its
potential energy. In accordance with Eqs.({\ref{epa}}) and
({\ref{kes}})
\begin{equation}
\frac{9}{14}\frac{G\mathbb{M}_\star^2}{\mathbb{R}_\star}=\frac{30}{7}k\mathbb{T}_\star
\mathbb{N}_\star.
\end{equation}
Introducing the non-dimensional parameter
\begin{equation}
\beta=\frac{G\mathbb{M}_\star \frac{A}{Z}m_p}{\mathbb{R}_\star
k\mathbb{T}_\star},\label{Bg}
\end{equation}
we obtain
\begin{equation}
\beta=\frac{20}{3}=6.67,\label{B}
\end{equation}
and at taking into account Eqs.({\ref{eta1}}) and ({\ref{tcore}}),
the core mass is
\begin{equation}
\mathbb{M}_\star=\left[\frac{20}{3}\left(\frac{25\cdot
13}{28}\right)^{1/3}\frac{3}{4\cdot
3.14}\right]^{3/2}\frac{\mathbb{M}_{Ch}}{{\left(\frac{A}{Z}\right)^2}}=6.84
\frac{\mathbb{M}_{Ch}}{{\left(\frac{A}{Z}\right)^2}}\label{Mcore}
\end{equation}
The obtained equation plays a very important role, because together
with Eq.(\ref{2mstar}), it gives a possibility to predict the total
mass of a star:
\begin{equation}
\mathbb{M}^{star}=2\mathbb{M}_\star=\frac{13.68
\mathbb{M}_{Ch}}{\left(\frac{A}{Z}\right)^2}\approx\frac{25.34
\mathbb{M}_\odot}{\left(\frac{A}{Z}\right)^2}.\label{M}
\end{equation}

The comparison of obtained prediction Eq.({\ref{M}}) with measuring
data gives a method to check our theory. Although there is no way to
determine  chemical composition of cores of far stars, some
predictions can be made in this way. At first, there must be no
stars which masses exceed the mass of the Sun by more than one and a
half orders, because it accords to limiting mass of stars consisting
from hydrogen with $A/Z=1$. Secondly, though the neutronization
process makes neutron-excess nuclei stable, there is no reason to
suppose that  stars with $A/Z>10$ (and with mass in hundred times
less than hydrogen stars) can exist. Thus, the theory predicts that
the whole mass spectrum must be placed in the interval from 0.25 up
to approximately 25 solar masses. These predications are verified by
measurements quite exactly. The mass distribution of binary
stars\footnote{The use of these data is caused by the fact that only
the measurement of parameters of binary star rotation gives a
possibility to determine their masses with satisfactory accuracy.}
is shown in Fig.{\ref{starM}} \cite{Heintz}.

\begin{figure}
\hspace{-3cm}
\includegraphics[scale=0.9]{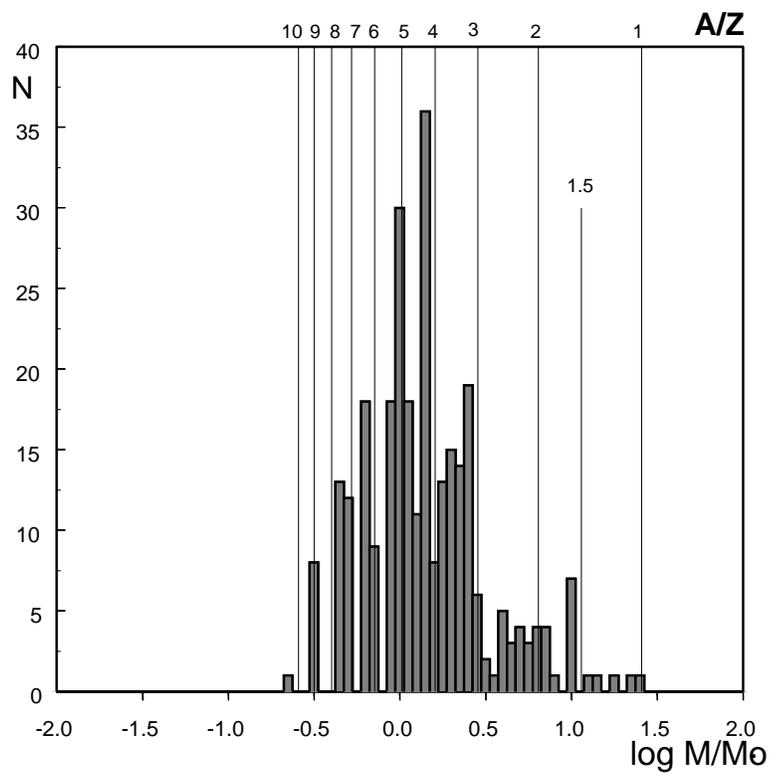}
\caption{The mass distribution of binary stars \cite{Heintz}. On
abscissa, the logarithm of the star mass over the Sun mass is shown.
Solid lines mark masses, which agree with selected values of A/Z
from Eq.(\ref{M}).}\label{starM}
\end{figure}

Besides, one can see the presence of separate peaks for stars with
$A/Z = 3; 4; 5...$ and with $A/Z = 3/2$ in Fig.{\ref{starM}}. At
that it is important to note, that according to Eq.(\ref{M}) the Sun
must consist from a substance with $A/Z=5$. This conclusion is in a
good agreement with results of consideration of solar oscillations
(Chapter \ref{Ch9}).
\begin{figure}
\hspace{-3cm}
\includegraphics[scale=0.9]{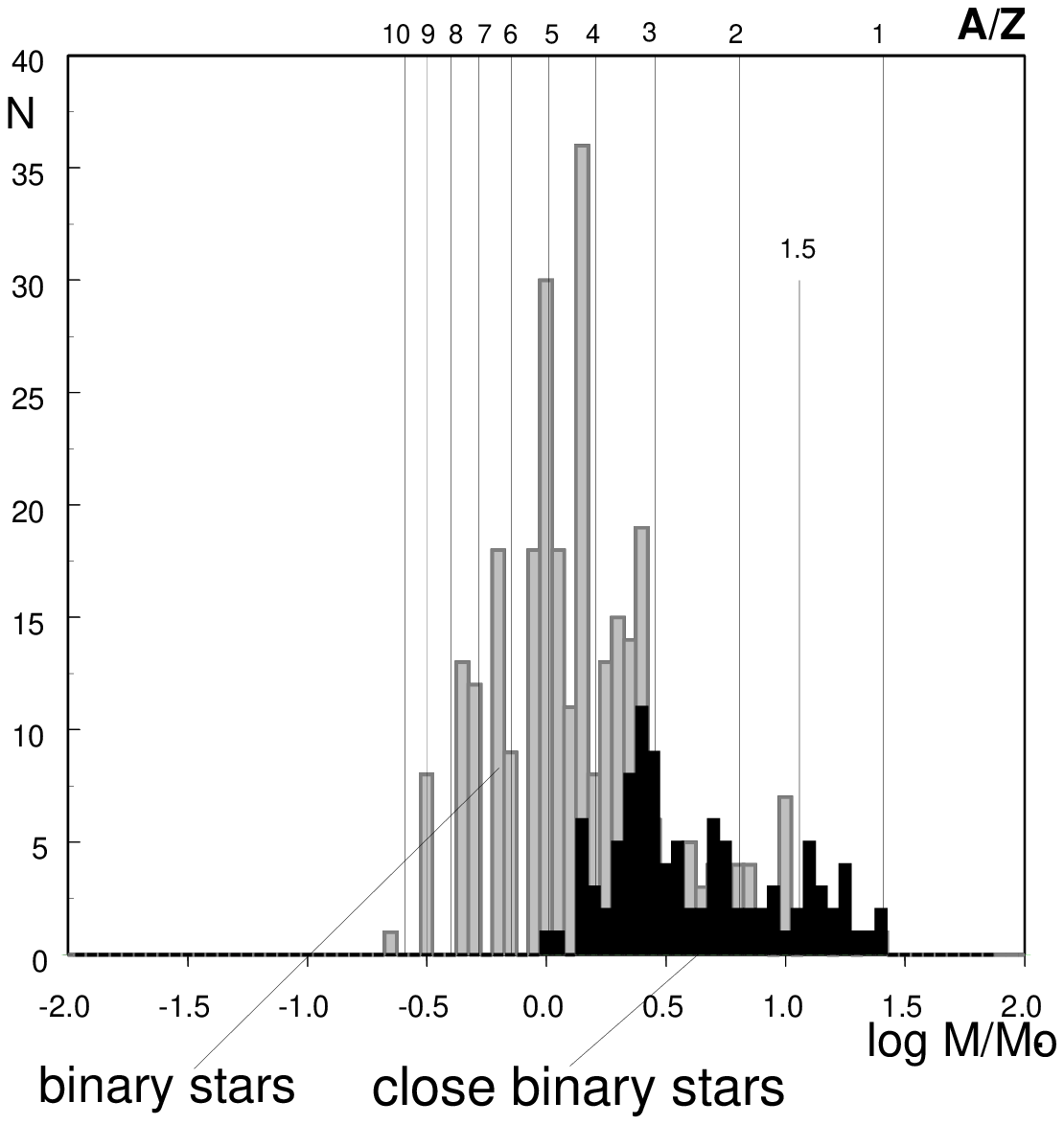}
\caption{The mass distribution of close binary stars \cite{Kh}. On
abscissa, the logarithm of the star mass over the Sun mass is shown.
Solid lines mark masses, which agree with selected values of A/Z
from Eq.(\ref{M}). The binary star spectrum is shown for
comparison.}\label{starM2}
\end{figure}
The mass spectrum of close binary stars\footnote{The data of these
measurements were obtained in different observatories of the world.
The last time the summary table with these data was gathered by
Khaliulilin Kh.F. (Sternberg Astronomical Institute) \cite{Kh} in
his dissertation (in Russian) which has unfortunately has a
restricted access. With his consent and for readers convenience, we
place that table in Appendix on http://astro07.narod.ru.}  is shown
in Fig.{\ref {starM2}}. It does not contain stars with high
parameter $A/Z$, but it is important that both spectra - those of
binary stars and those of close binary stars - come abruptly to the
end near $A/Z=1$.

\subsection{Radii of stars}
Using Eq.({\ref{eta1}}) and Eq.({\ref{Mcore}}), we can determine the
star core radius:
\begin{equation}
\mathbb{R}_\star=\frac{3(6.84\pi)^{1/3}}{4(Z+1)A/Z}{a_B}\left(\frac{\hbar
c}{G m_p^2}\right)^{1/2}  \approx \frac{1.41\cdot
10^{11}}{{(Z+1)A/Z}}cm.\label{Rcore}
\end{equation}
The temperature near the star surface is relatively small. It is
approximately  by 3 orders smaller than it is inside the core.
Because of it at calculation of surface parameters, we must take
into consideration effects of this order, i.e. it is necessary to
take into account the gravity action on the electron gas. At that it
is convenient to consider the plasma cell as some neutral quasi-atom
(like the Thomas-Fermi atom). Its electron shell is formed by a
cloud of free electrons.

Each such quasi-atom is retained on the star surface by its negative
potential energy
\begin{equation}
\left(\mathcal{E}_{gravitational}+\mathcal{E}_{electric}\right)<0.
\end{equation}
The electron cloud of the cell is placed in the volume $\delta
V=\frac{4\pi}{3}r_s^3$, (where
$r_s\approx\left(\frac{Z}{n_e}\right)^{1/3}$) under pressure $P_e$.
The evaporation of plasma cell releases energy $\mathcal{E}_{PV}=P_e
V_s$, and the balance equation takes the form:
\begin{equation}
\mathcal{E}_{gravitational}+\mathcal{E}_{electric}+\mathcal{E}_{PV}=0\label{srf}.
\end{equation}
In cold plasma, the electron cloud of the cell has energy
$\mathcal{E}_{PV}\approx{e^2}{n_e}^{1/3}$. in very hot plasma at
$kT\gg\frac{Z^2 e^2}{r_s}$, this energy is equal to
$\mathcal{E}_{PV}=\frac{3}{2}ZkT$. On the star surface these
energies are approximately equal:
\begin{equation}
\frac{k\mathbb{T}_0}{e^2 n_e^{1/3}}\approx
\frac{1}{\alpha}\left(\frac{\mathbb{R}_0}{\mathbb{R}_\star}\right)^2\approx
1.
\end{equation}
One can show it easily, that in this case
\begin{equation}
\mathcal{E}_{PV}\approx 2Z\sqrt{\frac{3}{2}kT\cdot{e^2}{n_e}^{1/3}}.
\end{equation}
And if to take into account Eqs.({\ref{an-r}})-({\ref{tr}}), we
obtain
\begin{equation}
\mathcal{E}_{PV}\approx 1.5 Z
k\mathbb{T}_\star\left(\frac{\mathbb{R}_\star}{\mathbb{R}_0}\right)^3\sqrt{\alpha\pi}
\end{equation}
The energy of interaction of a nucleus with its electron cloud does
not change at evaporation of the cell and it can be neglected. Thus,
for the surface
\begin{equation}
\mathcal{E}_{electric}=\frac{2\pi\mathfrak{P}^2}{3n_s}=\frac{2G\mathbb{M}_\star}{\mathbb{R}_0}
\left({A}m_p-{Z}m_e\right).
\end{equation}
The gravitational energy of the cell on the surface
\begin{equation}
\mathcal{E}_{gravitational}=-\frac{2G\mathbb{M}_\star}{\mathbb{R}_0}\left({A}m_p+{Z}m_e\right).
\end{equation}
Thus, the balance condition Eq.({\ref{srf}}) on the star surface
obtains the form
\begin{equation}
-\frac{4G\mathbb{M_\star}Zm_e}{\mathbb{R}_0}+ 1.5Z
k\mathbb{T}_\star\left(\frac{\mathbb{R}_\star}{\mathbb{R}_0}\right)^3\sqrt{\alpha\pi}=0.
\end{equation}
\subsection{The ${\mathbb{R}_\star}/{\mathbb{R}_0}$ ratio and ${\mathbb{R}_0}$}
With account of Eq.({\ref{tr}) and Eqs.({\ref{B})-({\ref{Bg}), we
can write
\begin{equation}
\frac{\mathbb{R}_0}{\mathbb{R}_\star}=\left(\frac{\sqrt{\alpha\pi}}{2\beta}\frac{\frac{A}{Z}m_p}{m_e}\right)^{1/2}
\approx {4.56}{\sqrt{\frac{A}{Z}}}\label{RRs}
\end{equation}
As the star core radius is known Eq.({\ref{Rcore}}), we can obtain
the star surface radius:
\begin{equation}
\mathbb{R}_0\approx\frac{6.44\cdot 10^{11}}{(Z+1)(A/Z)^{1/2}}
cm\label{R0}
\end{equation}

\subsection{The temperature of a star surface}
At known Eq.({\ref{tr}}) and Eq.({\ref{tcore}}), we can calculate
the temperature of external surface of a star
\begin{equation}
\mathbb{T}_0=\mathbb{T}_\star
\left(\frac{\mathbb{R}_\star}{\mathbb{R}_0}\right)^4 \approx
4.92\cdot 10^5 \frac{(Z+1)}{(A/Z)^2}\label{T0}
\end{equation}

\subsection{The comparison with measuring data}
The solar oscillations (see Chapter({\ref{Ch9}})) show that the Sun
consists basically from helium-10 with  $Z=2,A/Z=5$. With account of
this
\begin{equation}
\mathbb{R}_0({Z=2,A/Z=5})=9.59 \cdot 10^{10} cm,
\end{equation}
it differs from the measured value of the Sun radius
\begin{equation}
\mathbb{R}_\odot=6.96 \cdot 10^{10} cm.
\end{equation}
It can be a consequence of the estimation of the core radius from
Eq.({\ref{Rcore}}) at $Z=2,A/Z=5$ gives
$\mathbb{R}_\star(Z=2,A/Z=5)=9.6\cdot 10^9 cm$, and the ratio of
measured radius of the Sun to the calculated value of the core
radius is equal
\begin{equation}
\frac{R_\odot}{\mathbb{R}_\star(Z=2,A/Z=5)}\approx 7.24,\label{RRt}
\end{equation}
where as the calculation Eq.({\ref{RRs}}) gives
\begin{equation}
\frac{\mathbb{R}_0}{\mathbb{R}_\star(Z=2,A/Z=5)}\approx
10.2\label{RRt}
\end{equation}
At the same time, the calculated value of surface temperature of the
Sun (Eq.({\ref{T0}})),
\begin{equation}
\mathbb{T}_0\approx 5911 \, K
\end{equation}
is in a good agreement with its measured value
($\mathbb{T}_\odot\approx 5850 \, K$).

The calculation shows that the mass of core of the Sun
\begin{equation}
\mathbb{M}_\star(Z=2,A/Z=5)\approx 9.68\cdot 10^{32}~g\label{MtS}
\end{equation}
i.t. almost exactly equals to one half of full mass of the Sun
\begin{equation}
\frac{\mathbb{M}_\star(Z=2,A/Z=5)}{\mathbb{M}_\odot}\approx
0.486\label{MrS}
\end{equation}
in full agreement with Eq.({\ref{2mstar}}).

\vspace{1cm}

In addition to obtained determinations of the mass of a star
Eq.({\ref{M}}), its temperature  Eq.({\ref{T0}}) and its radius
Eq.({\ref{R0}}) give possibility
 to check the calculation, if we compare these results with measuring
 data.  Really, dependencies measured by astronomers  can be
 described by functions:
\begin{equation}
\mathbb{M}=\frac{Const_1}{(A/Z)^2},
\end{equation}
\begin{equation}
{\mathbb{R}}_0=\frac{Const_2}{(Z+1)(A/Z)^{1/2}},
\end{equation}
\begin{equation}
{\mathbb{T}}_0=\frac{Const_3(Z+1)}{(A/Z)^{2}}.
\end{equation}
If to combine they in the way, to exclude unknown parameter $(Z+1)$,
one can obtain relation:
\begin{equation}
{\mathbb{T}}_0 \mathbb{R}_0=Const\, \mathbb{M}^{5/4},\label{5/4}
\end{equation}
Its accuracy can by checked. For this checking, let us use the
measuring data of parameters of masses, temperatures and radii of
close binary stars \cite{Kh}.  The results of these measurements are
shown in Fig.({\ref{RT-M}}), where the dependence according to
Eq.({\ref{5/4}}). It is not difficult to see that these data are
well described by the calculated dependence. It speaks about
successfulness of our consideration.
\begin{figure}
\begin{center}
\includegraphics[scale=0.7]{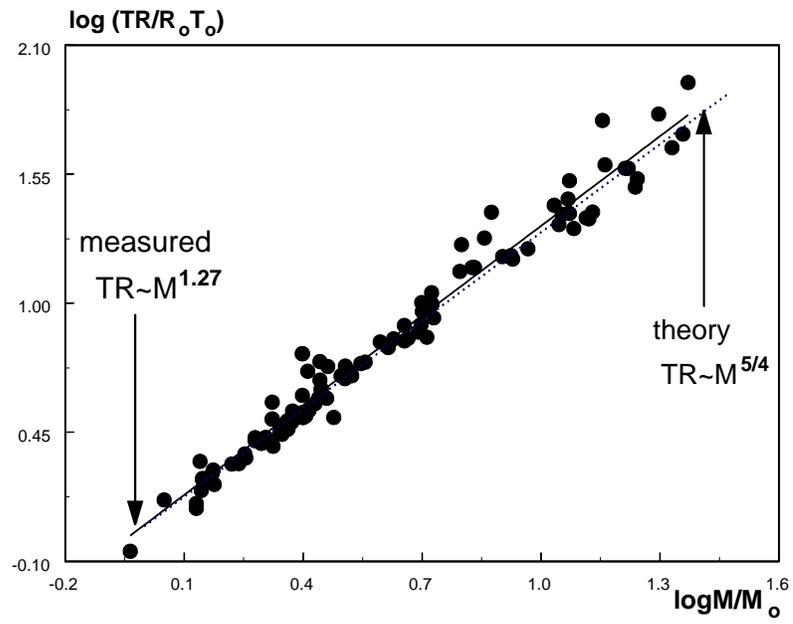}
\caption{The relation between main parameters of stars
(Eq.({\ref{5/4}})) and corresponding data of astronomical
measurements for close binary stars \cite{Kh} are shown.}
{\label{RT-M}}
\end{center}
\end{figure}

\chapter[The thermodynamic relations]{The thermodynamic relations of intrastellar plasma}
\label{Ch6}

\section[The thermodynamic relations]{The thermodynamic relation of star atmosphere parameters}

Hot stars steadily generate energy and radiate it from their
surfaces. This  is non-equilibrium radiation in relation to a star.
But it may be a stationary radiation for a  star in steady state.
Under this condition, the star substance can be considered as an
equilibrium. This condition can be considered as quasi-adiabatic,
because the interchange of energy between both subsystems -
radiation and substance - is stationary and it does not result in a
change of a steady state of substance. Therefore at consideration of
state of a star atmosphere, one can base it on equilibrium
conditions of hot plasma and the ideal gas law for adiabatic
condition can be used for it in the first approximation.

It is known, that thermodynamics can help to establish correlation
between  steady-state parameters of a system. Usually, the
thermodynamics considers systems at an equilibrium state with
constant temperature,  constant particle density and constant
pressure over all system. The characteristic feature of the
considered system is the existence of equilibrium at the absence of
a constant temperature and particle density over atmosphere of a
star. To solve this problem, one can introduce averaged pressure
\begin{equation}
\widehat{P}\approx\frac{G\mathbb{M}^2}{\mathbb{R}_0^4}\label{Pav},
\end{equation}
averaged temperature
\begin{equation}
\widehat{T}=\frac{\int_\mathbb{V} T dV}{V}\sim \mathbb{T}_0
\biggl(\frac{\mathbb{R}_0}{\mathbb{R}_\star}\biggr)\label{TR3}
\end{equation}
and averaged particle density
\begin{equation}
\widehat{n}\approx \frac{\mathbb{N}_A}{\mathbb{R}_0^3}
\end{equation}
After it by means of thermodynamical methods, one can find relation
between parameters of a star.

\subsection{The ${c_P}/{c_V}$ ratio}
At a movement of particles according to the theorem of the
equidistribution, the energy $kT/2$ falls at each degree of freedom.
It gives the heat capacity $c_v=3/2k$.

According to the virial theorem \cite{LL,VL}, the full energy of a
star should be equal to its kinetic energy (with opposite
sign)(Eq.({\ref{ek})), so as full energy related to one particle
\begin{equation}
\mathcal{E}=-\frac{3}{2}kT
\end{equation}
In this case the heat capacity at constant volume (per particle over
Boltzman constant $k$) by definition is
\begin{equation}
c_V=\biggl(\frac{dE}{dT}\biggr)_V=-\frac{3}{2}\label{cv}
\end{equation}
The negative heat capacity of stellar substance is not surprising.
It is a known fact and it is discussed in Landau-Lifshitz course
\cite{LL}. The own heat capacity of each particle of star substance
is positive. One obtains the negative heat capacity  at taking into
account the gravitational interaction between particles.

By definition the heat capacity of an ideal gas particle at
permanent pressure \cite{LL} is
\begin{equation}
c_P=\biggl(\frac{dW}{dT}\biggr)_P,\label{cp}
\end{equation}
where  $W$ is enthalpy of a gas.

As  for the ideal gas \cite{LL}
\begin{equation}
W-\mathcal{E}=NkT,
\end{equation}
and the difference between $c_P$ and $c_V$
\begin{equation}
{c_P-c_V}=1.
\end{equation}

Thus in the case considered, we have
\begin{equation}
c_P=-\frac{1}{2}.
\end{equation}
Supposing that conditions  are close to adiabatic ones, we can use
the equation of the Poisson's adiabat.

\subsection{The Poisson's adiabat}
The thermodynamical potential of a system consisting of $N$
molecules of ideal gas at temperature  $T$ and pressure $P$ can be
written as \cite{LL}:
\begin{equation}
\Phi=const\cdot N +NTlnP-Nc_P T lnT.
\end{equation}
The entropy of this system
\begin{equation}
S=const\cdot N -NlnP+Nc_P lnT.
\end{equation}
As at adiabatic process, the entropy remains constant
\begin{equation}
-NTlnP+Nc_P T lnT=const,
\end{equation}
we can write the equation for relation of averaged pressure in a
system with its volume (The Poisson's adiabat) \cite{LL}:
\begin{equation}
\widehat{P}V^{\widetilde{\gamma}}=const\label{Pua},
\end{equation}
where $\widetilde{\gamma}=\frac{c_P}{c_V}$ is the exponent of
 adiabatic constant. In considered case taking into account of
Eqs.({\ref{cp}}) and ({\ref{cv}}), we obtain
\begin{equation}
\widetilde{\gamma}=\frac{c_P}{c_V}=\frac{1}{3}.
\end{equation}
As $V^{1/3}\sim \mathbb{R}_0$, we have for equilibrium condition
\begin{equation}
\widehat{P}\mathbb{R}_0=const\label{Pub}.
\end{equation}

\section{The mass-radius ratio}
Using Eq.({\ref{Pav}}) from  Eq.({\ref{Pub}}), we use the equation
for dependence of masses of stars on their radii:
\begin{equation}
\frac{\mathbb{M}^2}{\mathbb{R}_0^3}=const\label{rm23}
\end{equation}
This equation shows the existence of internal constraint of chemical
parameters of equilibrium state of a star. Indeed, the substitution
of obtained determinations Eq.({\ref{R0}}) and ({\ref{T0})) into
Eq.({\ref{rm23}}) gives:
\begin{equation}
(1+Z)\sim (A/Z)^{5/6}\label{Z-AZ}
\end{equation}
Simultaneously the observational data of  masses, of radii and their
temperatures was obtained by astronomers for close binary stars
\cite{Kh}. The dependence of radii of these stars over these masses
is shown in Fig.{\ref{RM}} on double logarithmic scale. The solid
line shows the result of fitting of measurement data
$\mathbb{R}_0\sim \mathbb{M}^{0.68}$. It is close to theoretical
dependence $\mathbb{R}_0\sim \mathbb{M}^{2/3}$~(Eq.{\ref{rm23}})
which is shown by dotted line.

\begin{figure}
\begin{center}
\includegraphics[scale=0.7]{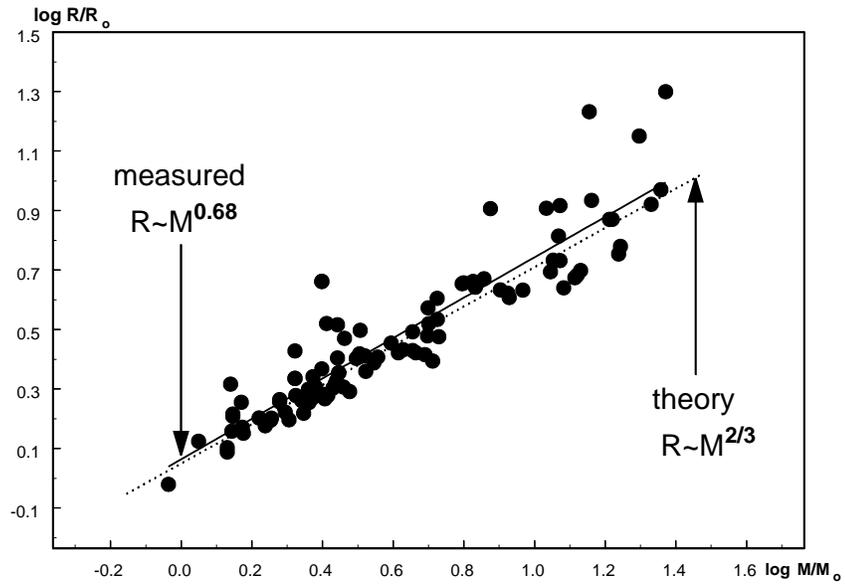}
\caption{The dependence of  radii of stars over the star mass
\cite{Kh}. Here the radius of stars is normalized to the sunny
radius, the stars masses are normalized to the mass of the Sum. The
data are shown on double logarithmic scale. The solid line shows the
result of fitting of measurement data $\mathbb{R}_0\sim
\mathbb{M}^{0.68}$. The theoretical dependence
$\mathbb{R}_0\sim\mathbb{M}^{2/3}$~({\ref{rm23}}) is shown by the
dotted line.}\label{RM}
\end{center}
\end{figure}

\section[The mass-temperature-luminosity relations]{The mass-temperature and mass-luminosity relations}

Taking into account Eqs.({\ref{tr}}), ({\ref{Ts1}}) and ({\ref{RN}})
one can obtain the relation between surface temperature and the
radius of a star
\begin{equation}
\mathbb{T}_0\sim \mathbb{R}_0^{7/8},
\end{equation}
or accounting for  ({\ref{rm23}})
\begin{equation}
\mathbb{T}_0\sim \mathbb{M}^{7/12}\label{tm}
\end{equation}
The dependence of the temperature on the star surface over the star
mass of close binary stars \cite{Kh} is shown in Fig.(\ref{TM}).
Here the temperatures of stars are normalized to the sunny surface
temperature (5875~C), the stars masses are normalized to the mass of
the Sum. The data are shown on double logarithmic scale. The solid
line shows the result of fitting of measurement data
($\mathbb{T}_0\sim \mathbb{M}^{0.59}$). The theoretical dependence
$\mathbb{T}_0\sim
\mathbb{M}^{7/12}$~(Eq.{\ref{tm}}) is shown by dotted line.} 

\begin{figure}
\begin{center}
\includegraphics[scale=0.7]{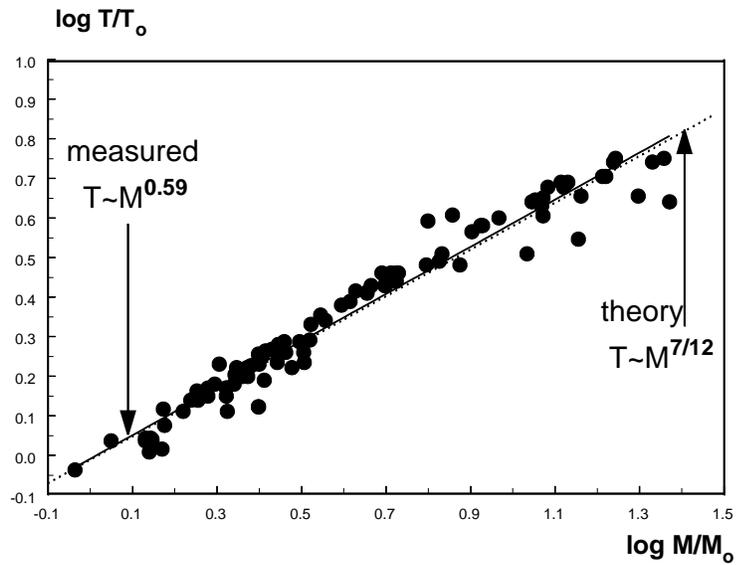}
\caption{The dependence of the temperature on the star surface over
the star mass of close binary stars \cite{Kh}. Here the temperatures
of stars are normalized to surface temperature of the Sun (5875~C),
the stars masses are normalized to the mass of Sum. The data are
shown on double logarithmic scale. The solid line shows the result
of fitting of measurement data ($\mathbb{T}_0\sim
\mathbb{M}^{0.59}$). The theoretical dependence $\mathbb{T}_0\sim
\mathbb{M}^{7/12}$~(Eq.{\ref{tm}}) is shown by dotted line.}
\label{TM}
\end{center}
\end{figure}
The luminosity of a star
\begin{equation}
\mathbb{L}_0\sim \mathbb{R}_0^2 \mathbb{T}_0^4.
\end{equation}
at taking into account (Eq.{\ref{rm23}}) and (Eq.{\ref{tm}}) can be
expressed as
\begin{equation}
\mathbb{L}_0\sim \mathbb{M}^{11/3}\sim \mathbb{M}^{3.67}\label{lm}
\end{equation}
This dependence is shown in Fig.(\ref{LM})
\begin{figure}
\begin{center}
\includegraphics[scale=0.7]{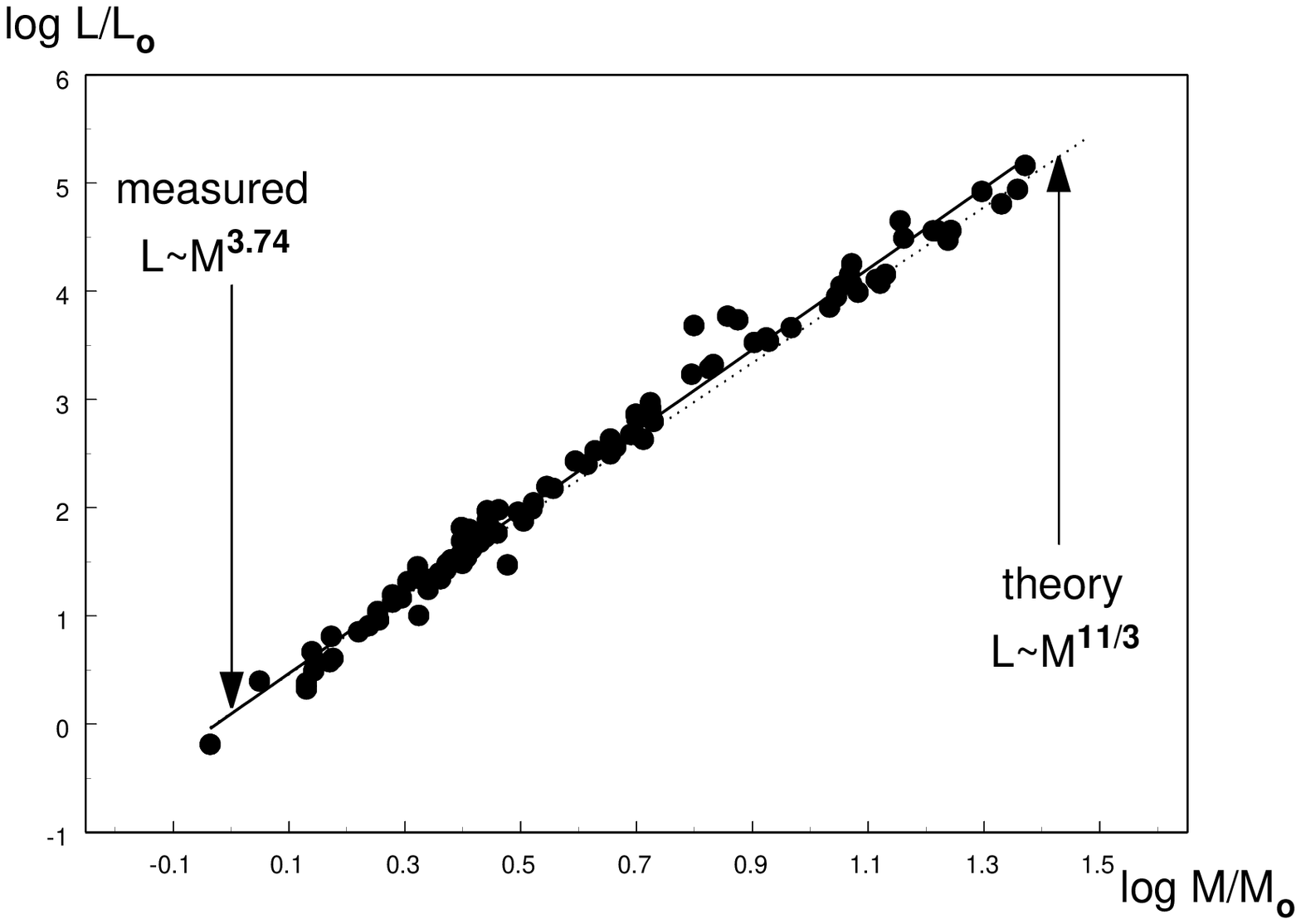}
\caption{The dependence of star luminosity on the star mass of close
binary stars \cite{Kh}. The luminosities are normalized to the
luminosity of the Sun, the stars masses are normalized to the mass
of the Sum. The data are shown on double logarithmic scale. The
solid line shows the result of fitting of measurement data
$\mathbb{L}\sim \mathbb{M}^{3.74} $. The theoretical dependence
$\mathbb{L}\sim \mathbb{M}^{11/3}$~(Eq.{\ref{lm}}) is shown by
dotted line.} \label{LM}
\end{center}
\end{figure}
It can be seen that all calculated interdependencies
$\mathbb{R(\mathbb{M})}$,$\mathbb{T(\mathbb{M})}$ and
$\mathrm{L(\mathbb{M})}$ show a good qualitative agreement with the
measuring data. At that it is important, that the quantitative
explanation of mass-luminosity dependence discovered at the
beginning of 20th century is obtained.

\chapter[Magnetic fields of stars]
{Magnetic fields and magnetic moments of stars} \label{Ch7}

\section[Magnetic moments of stars]{Magnetic moments of celestial bodies}

A thin spherical surface with radius  $r$ carrying an electric
charge $q$ at the rotation around its axis with frequency  $\Omega$
obtains the magnetic moment
\begin{equation}
{{\boldsymbol{\mathfrak{m}}}}=\frac{ r^2}{3c}q\boldsymbol\Omega.
\end{equation}
The rotation of a ball charged at density $\varrho(r)$ will induce
the magnetic moment
\begin{equation}
\boldsymbol{\mathfrak{\mu}}=\frac{\boldsymbol\Omega}{3c}\int_0^R
r^2\varrho(r) ~ 4\pi r^2dr.
\end{equation}
Thus the positively charged core of a star induces the magnetic
moment
\begin{equation}
\boldsymbol{\mathfrak{m}}_+=\frac{\sqrt{G}\mathbb{M}_\star
\mathbb{R}_\star^2}{5c}\boldsymbol\Omega.
\end{equation}
A negative charge will be concentrated in the star atmosphere. The
absolute value of atmospheric charge is equal to the positive charge
of a core. As the atmospheric charge is placed near the surface of a
star, its magnetic moment will be more than the core magnetic
moment. The calculation shows that as a result, the total magnetic
moment of the star will have the same order of magnitude as the core
but it will be negative:
\begin{equation}
\boldsymbol{\mathfrak{m}}_\Sigma\approx
-\frac{\sqrt{G}}{c}\mathbb{M}_\star
\mathbb{R}_\star^2\boldsymbol\Omega.
\end{equation}
Simultaneously, the torque of a ball with mass $\mathbb{M}$ and
radius  $\mathbb{R}$ is
\begin{equation}
\boldsymbol{\mathcal{L}}\approx \mathbb{M}_\star
\mathbb{R}_\star^2\boldsymbol\Omega.
\end{equation}
As a result, for celestial bodies where the force of their gravity
induces the electric polarization according to Eq.({\ref{roe}}), the
giromagnetic ratio will depend on world constants only:
\begin{equation}
\frac{{\boldsymbol{\mathfrak{m}}}_\Sigma}{\boldsymbol{\mathcal{L}}}\approx
-\frac{\sqrt{G}}{c}\label{ML}.
\end{equation}
This relation was obtained for the first time by P.M.S.Blackett
\cite{Blackett}. He shows that giromagnetic ratios of the Earth, the
Sun and the star 78 Vir are really near to $\sqrt{G}/c$.

By now the magnetic fields, masses, radii and velocities of rotation
are known for all planets of the Solar system and for a some stars
\cite{Sirag}. These measuring data are shown in Fig.({\ref{black}}),
which is taken from \cite{Sirag}. It is possible to see that these
data are in satisfactory agreement with Blackett's ratio.
\begin{figure}
\begin{center}
\includegraphics[scale=0.7]{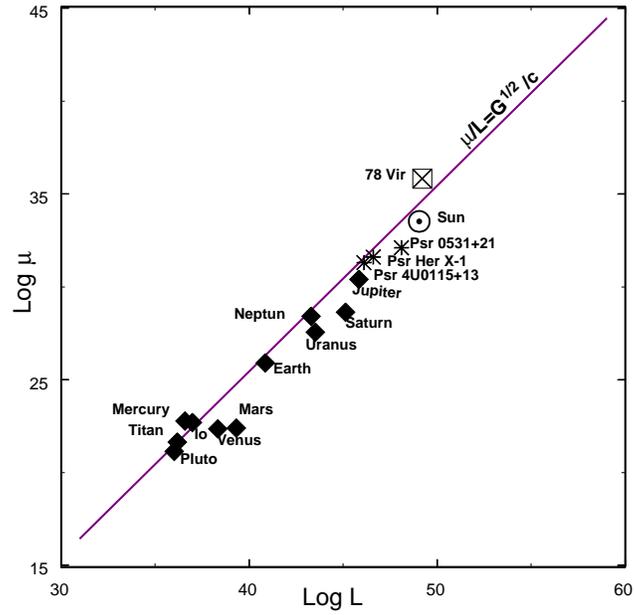}
\vspace{6cm} \caption {The observed values of magnetic moments of
celestial bodies vs. their angular momenta \cite{Sirag}. In
ordinate, the logarithm of the magnetic moment (in $Gs\cdot{cm^3}$)
is plotted; in abscissa the logarithm of the angular momentum (in
$erg\cdot{s}$) is shown. The solid line illustrates Eq.({\ref{ML}}).
The dash-dotted line fits of observed values.} \label{black}
\end{center}
\end{figure}
At some assumption, the same parameters can be calculated for
pulsars. All measured masses of pulsars are equal by the order of
magnitude \cite{Thor}. It is in satisfactory agreement with the
condition of equilibrium of relativistic matter Eq.({\ref{m-puls}}).
It gives a possibility to consider that masses and radii of pulsars
are determined. According to generally accepted point of view,
pulsar radiation is related with its rotation, and it gives their
rotation velocity. These assumptions permit to calculate the
giromagnetic ratios for three pulsars with known magnetic fields on
their poles \cite{Beskin}. It is possible to see from
Fig.({\ref{black}}), the giromagnetic ratios of these pulsars are in
agreement with Blackett's ratio.

\section[Magnetic fields of stars]{Magnetic fields of hot stars}
To make  the above rough estimation for magnetic fields induced by
the star atmosphere more accurate, we can take into account the
distribution of electric charges inside the star atmosphere
In accordance with this distribution, the atmosphere at its rotation
induces the moment
\begin{equation}
{\boldsymbol{\mathfrak{m}}}_-=-\frac{\sqrt{G}\mathbb{M}_\star\boldsymbol\Omega{\mathbb{R}_\star}^2}{9c}\int_{\mathbb{R}_\star}^{R_0}
r^2\frac{d}{dr}\sqrt{\xi^3(2-\xi^3)}~dr,
\end{equation}
Where  $\xi=\frac{\mathbb{R_\star}}{r}$.

As $\mathbb{R}_\star\ll R_0$, the magnetic moment of a core can be
neglected. In this case, the total magnetic moment of a star
\begin{equation}
{\boldsymbol{\mathfrak{m}}}\approx
-2\frac{\sqrt{G}\mathbb{M_\star}}{3c}\left(\frac{R_0^{7/2}}{\mathbb{R}_\star^{3/2}}\right)\boldsymbol\Omega.
\end{equation}
and magnetic field near the pole of a star
\begin{equation}
\boldsymbol{\mathcal{H}}\approx
-\frac{2\sqrt{G}\mathbb{M_\star}}{9c}\frac{R_0^{1/2}}{\mathbb{R}_\star^{3/2}}\boldsymbol{\Omega}.
\end{equation}
Using of substitution of  relations obtained above, one can conclude
that the magnetic field on a star pole must not depend on the
parameter $(Z+1)$ and very slightly depends on $A/Z$, and this
dependence can be neglected. I.e. the calculations show that this
field practically is not depending on the mass, on the radius and on
the temperature of a hot star and must depend on the velocity of
star rotation only:
\begin{equation}
{\boldsymbol{\mathcal{H}}}\approx
-0.47\left(\frac{m_e}{m_p}\right)^{3/4}\frac{\alpha
c}{{G}^{1/2}\beta^{1/4}}{\boldsymbol\Omega}\approx -8.8\cdot10^8
\boldsymbol{\Omega} {\quad}Oe.\label{Ht}
\end{equation}
The magnetic fields are measured for stars of Ap-class \cite{rom}.
These stars are characterized by changing their brightness in time.
The periods of these changes are measured too. At present there  is
no full understanding of causes of these visible changes of the
luminosity. If these luminosity changes caused by some internal
reasons will occur not uniformly on a star surface, one can conclude
that the measured period of the luminosity change can depend on star
rotation. It is possible to think that at relatively rapid rotation
of a star, the period of a visible change of the luminosity can be
determined by this rotation in general. To check this suggestion, we
can compare the calculated dependence (Eq.{\ref{Ht}}) with measuring
data \cite{rom} (see Fig. {\ref{H-W}}). It can be seen  that if
magnetic fields of hot stars are really described by the  mechanism
considered, there are unaccounted factors in this model.
\begin{figure}
\begin{center}
\includegraphics[scale=0.7]{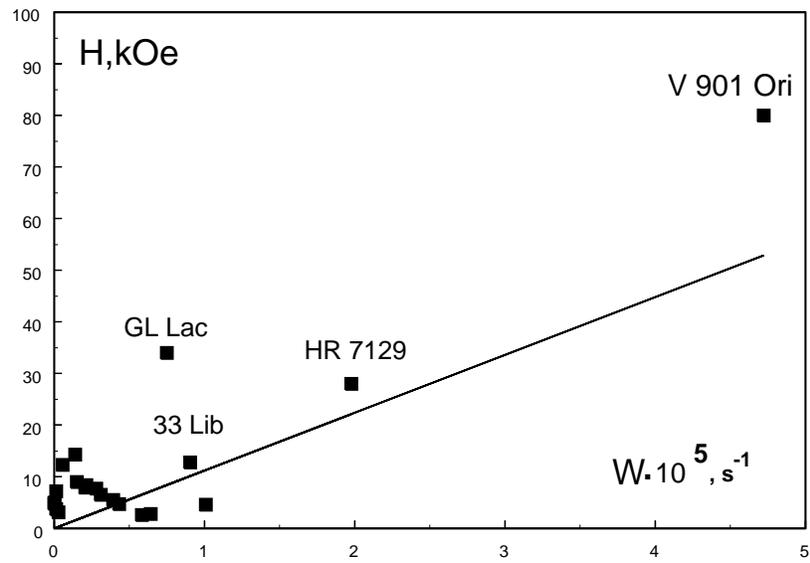}
\vspace{6cm} \caption {The dependence of magnetic fields on poles of
Ap-stars as a function of  their rotation velocity \cite{rom}. The
line shows Eq.(\ref{Ht})).}\label{H-W}
\end{center}
\end{figure}
It should be said that Eq.(\ref{Ht}) does not working well in case
with the Sun. The Sun surface rotates with period $T\approx 25\div
30$ days. At this velocity of rotation, the magnetic field on the
Sun pole calculated accordingly to Eq.(\ref{Ht}) must be about 1
kOe. The dipole field of Sun according to experts estimation is
approximately 20 times lower. There can be several reasons for that.

\chapter[The apsidal rotation]
{The angular velocity of the apsidal rotation in binary stars}
\label{Ch8}

\section[The apsidal rotation]{The apsidal rotation of close binary stars}
The apsidal rotation (or periastron rotation) of close binary stars
is a result of their non-Keplerian movement which originates from
the non-spherical form of stars. This non-sphericity has been
produced by rotation of stars around their axes or by their mutual
tidal effect. The second effect is usually smaller  and can be
neglected. The first and basic theory of this effect was developed
by A.Clairault at the beginning of the XVIII century. Now this
effect was measured for approximately 50 double stars. According to
Clairault's theory the velocity of periastron rotation must be
approximately 100 times faster if matter is uniformly distributed
inside a star. Reversely, it would be absent if all star mass is
concentrated in the star center. To reach an agreement between the
measurement data and calculations, it is necessary to assume that
the density of substance grows in direction to the center of a star
and  here it runs up to a value which is hundreds times greater than
mean density of a star. Just the same mass concentration of the
stellar substance is supposed by all standard theories of a star
interior. It has been usually considered as a proof of astrophysical
models. But it can be considered as a qualitative argument. To
obtain a quantitative agreement between theory and measurements, it
is necessary to fit parameters of the stellar substance distribution
in each case separately.

Let us consider this problem with taking into account the gravity
induced electric polarization of plasma in a star. As it was shown
above, one half of full mass of a star is concentrated in its plasma
core at a permanent density. Therefor, the effect of periastron
rotation of close binary stars must be reviewed with the account of
a change of forms of these star cores.

According to \cite{peri2},\cite{peri1} the ratio of the angular
velocity $\omega$ of rotation of periastron which is produced by the
rotation of a star around its axis with  the angular velocity
$\Omega$ is
\begin{equation}
\frac{\omega}{\Omega}=\frac{3}{2}\frac{(I_A-I_C)}{Ma^2}
\end{equation}
where $I_A$ and $I_C$ are the moments of inertia relatively to
principal axes of the ellipsoid. Their difference is
\begin{equation}
I_A-I_C=\frac{M}{5}(a^2-c^2),
\end{equation}
where $a$ and $c$ are the equatorial and polar radii of the star.

Thus we have
\begin{equation}
\frac{\omega}{\Omega}\approx \frac{3}{10}\frac{(a^2-c^2)}{a^2}.
\end{equation}

\section[The equilibrium form of the core]{The equilibrium form of the core of a rotating star}
In the absence of rotation the equilibrium equation  of plasma
inside star core (Eq.{\ref{Eu2}} is
\begin{equation}
\gamma {\bf g}_G+\rho_G {\bf E}_G=0\label{qm}
\end{equation}
where $\gamma$,${\bf g}_G$, $\rho_G$ and ${\bf E}_G$ are the
substance density the acceleration of gravitation, gravity-induced
density of charge and intensity of gravity-induced electric field
($div~{\bf g}_G=4\pi~ G~ \gamma$, $div~{\bf E}_G=4\pi \rho_G$ and
$\rho_G=\sqrt{G}\gamma$).

One can suppose, that at rotation,  under action of a rotational
acceleration  ${\bf g}_\Omega$, an additional electric charge with
density $\rho_\Omega$ and electric field ${\bf E}_\Omega$ can exist,
and the equilibrium equation obtains the form:

\begin{equation}
(\gamma_G+\gamma_\Omega)({\bf g}_G+{\bf
g}_\Omega)=(\rho_G+\rho_\Omega)({\bf E}_G+{\bf E}_\Omega),
\end{equation}

where

\begin{equation}
div~({\bf E}_G+{\bf E}_\Omega)=4\pi(\rho_G+\rho_\Omega)
\end{equation}

or

\begin{equation}
div~{\bf E}_\Omega=4\pi\rho_\Omega.
\end{equation}

We can look for a solution for electric potential in the form

\begin{equation}
\varphi=C_\Omega~r^2(3cos^2\theta-1)
\end{equation}

or in Cartesian coordinates

\begin{equation}
\varphi=C_\Omega(3z^2-x^2-y^2-z^2)
\end{equation}

where $C_\Omega$ is a constant.

 Thus

\begin{equation}
E_x=2~C_\Omega~x,~ E_y=2~C_\Omega~y,~ E_z=-4~C_\Omega~z
\end{equation}

and

\begin{equation}
div~{\bf E}_\Omega=0
\end{equation}

and we obtain important equations:

\begin{equation}
\rho_\Omega=0;
\end{equation}

\begin{equation}
\gamma g_\Omega=\rho {\bf E}_\Omega.
\end{equation}

Since centrifugal force must be contra-balanced by electric force

\begin{equation}
\gamma~2\Omega^2~x=\rho~2C_\Omega~x
\end{equation}

and

\begin{equation}
C_\Omega=\frac{\gamma~\Omega^2}{\rho}=\frac{\Omega^2}{\sqrt{G}}
\end{equation}

The potential of a positive uniform charged ball is

\begin{equation}
\varphi(r)=\frac{Q}{R}\biggl(\frac{3}{2}-\frac{r^2}{2R^2}\biggr)
\end{equation}

The negative  charge on the surface of a sphere induces inside the
sphere the potential

\begin{equation}
\varphi(R)=-\frac{Q}{R}
\end{equation}

where according to Eq.({\ref{qm}}) $Q=\sqrt{G}M$, and $M$ is the
mass of the star.

Thus the total potential inside the considered star is

\begin{equation}
\varphi_\Sigma=\frac{\sqrt{G}M}{2R}\biggl(1-\frac{r^2}{R^2}\biggr)+\frac{\Omega^2}{\sqrt{G}}r^2(3cos^2\theta-1)
\end{equation}

Since the  electric potential must be equal to zero on the surface
of the star, at $r=a$ and $r=c$

\begin{equation}
\varphi_\Sigma=0
\end{equation}

and  we obtain the equation which describes the equilibrium form of
the core of a rotating star (at $\frac{a^2-c^2}{a^2}\ll 1$)

\begin{equation}
\frac{a^2-c^2}{a^2}\approx\frac{9}{2\pi}\frac{\Omega^2}{G\gamma}\label{ef}.
\end{equation}

\section[The angular velocity]{The angular velocity of the apsidal rotation}

Taking into account of Eq.({\ref{ef}}) we have

\begin{equation}
\frac{\omega}{\Omega}\approx
\frac{27}{20\pi}\frac{\Omega^2}{G\gamma}\label{oo}
\end{equation}
If both stars of a close pair induce a rotation of periastron, this
equation transforms to

\begin{equation}
\frac{\omega}{\Omega}\approx
\frac{27}{20\pi}\frac{\Omega^2}{G}\biggl(\frac{1}{\gamma_1}+\frac{1}{\gamma_2}\biggr),
\end{equation}
where $\gamma_1$ and $\gamma_2$ are densities of star cores.

The equilibrium density of star cores is known (Eq.({\ref{eta1}})):

\begin{equation}
\gamma=\frac{16}{9\pi^2}\frac{A}{Z}m_p\frac{(Z+1)^3}{a_B^3}\label{go}.
\end{equation}

If we introduce  the period of ellipsoidal rotation
$P=\frac{2\pi}{\Omega}$ and  the period of the rotation of
periastron $U=\frac{2\pi}{\omega}$,  we obtain from Eq.({\ref{oo}})

\begin{equation}
\frac{\mathcal{P}}{\mathcal{U}}\biggl(\frac{\mathcal{P}}{\mathcal{T}}\biggr)^2\approx\sum_1^2\xi_i\label{2},
\end{equation}
where
\begin{equation}
\mathcal{T}=\sqrt{\frac{243~\pi^3}{80}}~\tau_0\approx 10 \tau_0,
\end{equation}
\begin{equation}
\tau_0=\sqrt{\frac{a_B^3}{G~m_p}}\approx 7.7\cdot 10^2 sec
\end{equation}
and
\begin{equation}
\xi_i=\frac{Z_i}{A_i(Z_i+1)^3}\label{2P}.
\end{equation}

\section[The comparison  with observations]{The comparison of the calculated angular velocity of the periastron rotation with observations}

Because the substance density (Eq.({\ref{go}})) is depending
approximately on the second power of the nuclear charge, the
periastron movement of stars consisting of heavy elements will fall
out from the observation as it is very slow. Practically the
obtained equation ({\ref{2}}) shows that it is possible to observe
the periastron rotation of a star consisting of light elements only.

The value $\xi=Z/[A(Z+1)^3]$ is equal to $1/8$ for hydrogen,
$0.0625$ for deuterium, $1.85\cdot 10^{-2}$ for helium. The
resulting value of the periastron rotation of double stars will be
the sum of separate stars rotation. The possible combinations of a
couple and their value of $\sum_1^2\xi_i$ for stars consisting of
light elements is shown in Table {\ref{peri}}.

\bigskip

\begin{tabular}{||c|c|c||}\hline\hline
  star1&star2 &$\xi_1+\xi_2$\\
  composed of &composed of&\\\hline
  H & H & .25\\
  H & D & 0.1875\\
  H & He & 0.143\\
  H & hn & 0.125\\
  D & D & 0.125 \\
  D & He & 0.0815 \\
  D & hn & 0.0625 \\
  He & He & 0.037 \\
  He & hn & 0.0185 \\ \hline\hline
\end{tabular}\label{peri}
Table {\ref{peri}}
\bigskip

The "hn" notation in Table {\ref{peri}} indicates that the second
component of the couple consists of heavy elements or it is a dwarf.

The results of measuring of main parameters for close binary stars
are gathered in \cite{Kh}. For reader convenience, the data of these
measurement is applied in the Table in Appendix. One can compare our
calculations with  data of these measurements. The distribution of
close binary stars on value of $(\mathcal{P}/\mathcal{U})
(\mathcal{P}/\mathcal{T})^2$ is shown in Fig.{\ref{periastr}} on
logarithmic scale. The lines mark the values of parameters
$\sum_1^2\xi_i$ for different light atoms in accordance  with
{\ref{2P}}. It can be seen that calculated values the periastron
rotation for stars composed by light elements which is summarized in
Table{\ref{peri}} are in  good agreement with separate peaks of
measured data. It confirms that our approach to interpretation of
this effect is adequate to produce  a satisfactory accuracy of
estimations.
\begin{figure}
\hspace{-1.5cm}
\includegraphics[scale=0.7]{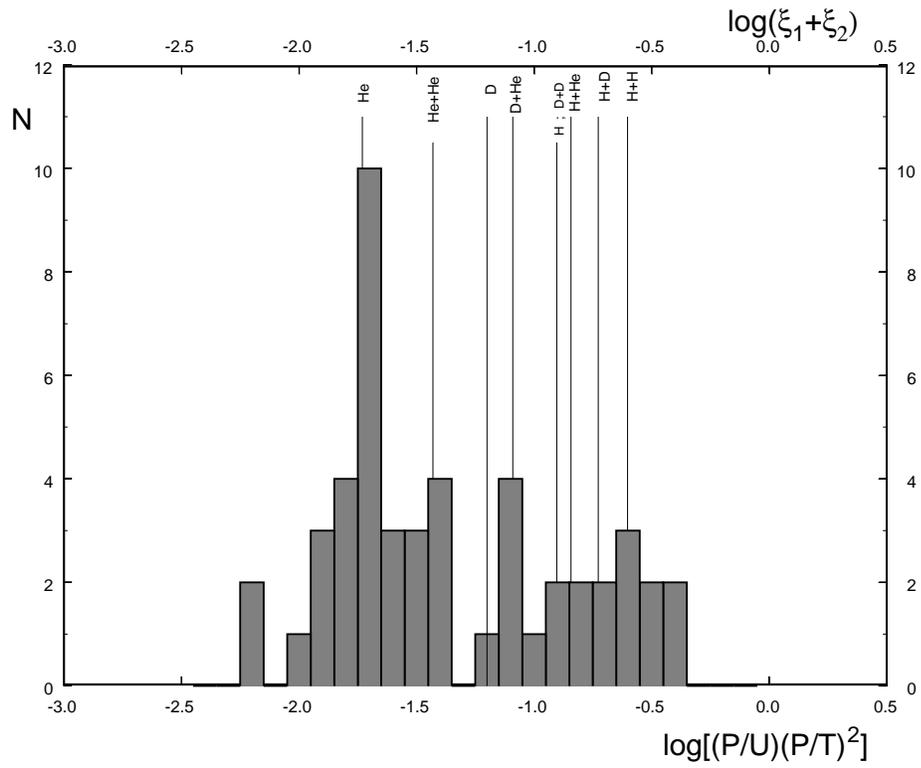}
\caption{The distribution of close binary stars \cite{Kh} on value
of $(\mathcal{P}/\mathcal{U})(\mathcal{P}/\mathcal{T})^2$. Lines
show  parameters $\sum_1^2\xi_i$ for different light atoms in
according with {\ref{2P}}.} \label{periastr}
\end{figure}

\chapter{The solar seismical oscillations}\label{Ch9}

\section [The spectrum  of solar oscillations]{The spectrum  of solar seismic oscillations}

The measurements \cite{bison} show that the Sun surface is subjected
to a seismic vibration. The most intensive oscillations have the
period about five minutes and the wave length about $10^4$km or
about hundredth part of the Sun radius. Their spectrum obtained by
BISON collaboration is shown in Fig.{\ref{bison}}.

\begin{figure}
\begin{center}
\includegraphics[scale=0.7]{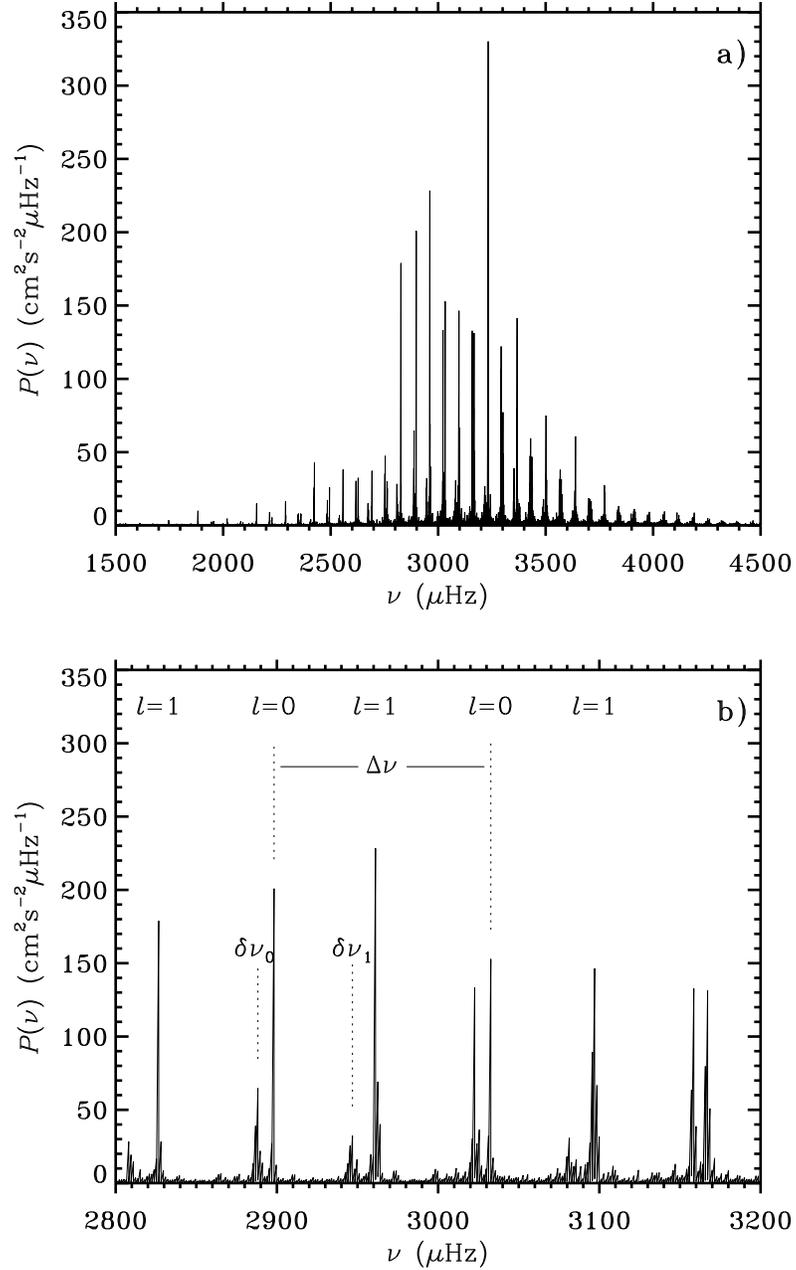}
\caption{
$(a)$ The power spectrum of solar oscillation obtained by
means of Doppler velocity measurement in light integrated over the
solar disk. The data were obtained from
the BISON network \cite{bison}. 
$(b)$ An expanded view of a part of frequency range.\label{bison}}
\end{center}
\end{figure}

It is supposed, that these oscillations are a superposition of a big
number of different modes of resonant acoustic vibrations, and that
acoustic waves propagate in different trajectories in the interior
of the Sun and they have multiple reflection from surface. With
these reflections trajectories of same waves can be closed and as a
result standing waves are forming.

Specific features of spherical body oscillations are described by
the expansion in series on spherical functions. These oscillations
can have a different number of wave lengths on the radius of a
sphere ($n$) and  a different number of wave lengths on its surface
which is determined by the $l$-th spherical harmonic. It is accepted
to describe the sunny surface oscillation spectrum as the expansion
in series \cite{CD}:
\begin{equation}
\nu_{nlm} \simeq \Delta \nu_0(n+\frac{l}{2}+\epsilon_0)-l(l+1)D_0 +
m\Delta \nu_{rot}.\label{nu}
\end{equation}
Where the last item is describing the effect of the Sun rotation and
is small. The main contribution is given by the first item which
creates a large splitting in the spectrum (Fig.{\ref{bison}})
\begin{equation}
\triangle\nu=\nu_{n+1,l}-\nu_{n,l}.
\end{equation}
The small splitting of spectrum (Fig.{\ref{bison}}) depends on the
difference
\begin{equation}
\delta\nu_l=\nu_{n,l}-\nu_{n-1,l+2}\approx (4l+6)D_0.
\end{equation}
A satisfactory agreement of these estimations and measurement data
can be obtained at \cite{CD}

\begin{equation}
\Delta \nu_0=120~\mu Hz,~ \epsilon_0=1.2,~ D_0=1.5~\mu Hz,~ \Delta
\nu_{rot}=1\mu Hz.\label{del}
\end{equation}

 To obtain these values of parameters $\Delta \nu_0,~ \epsilon_0
è ~D_0$ from theoretical models is not possible. There are a lot of
qualitative and quantitative assumptions used at a model
construction and a direct calculation of spectral frequencies
transforms into a unresolved complicated problem.

Thus, the current interpretation of the measuring spectrum by the
spherical harmonic  analysis does not make it clear. It gives no
hint to an answer to the question: why oscillations close to
hundredth harmonics are really excited and there are no waves near
fundamental harmonic?

The measured spectra have a very  high resolution (see
Fig.({\ref{bison}})). It means that an oscillating system has high
quality. At this condition, the system must have oscillation on a
fundamental frequency. Some peculiar mechanism must exist to force a
system to oscillate on a high harmonic. The current explanation does
not clarify it.

It is important, that now the solar oscillations are measured by
means of two different methods. The solar oscillation spectra which
was obtained on program "BISON", is shown on Fig.({\ref{bison}})).
It has a very high resolution, but (accordingly to the Liouville's
theorem) it was obtained with some loss of luminosity, and as a
result not all lines are well statistically worked.
\begin{figure}
\hspace{.5cm}
\includegraphics[scale=1]{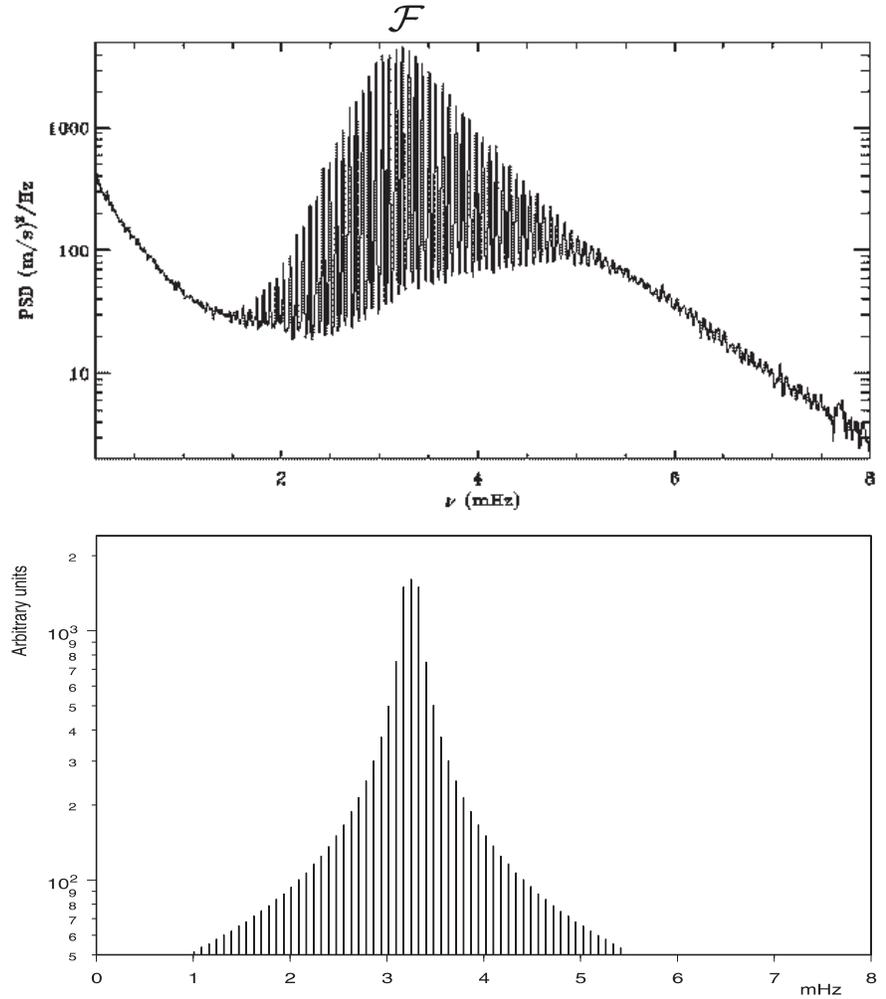}
\caption{$(a)$ The measured power spectrum of solar oscillation. The
data were obtained from the SOHO/GOLF measurement \cite{soho}. $(b)$
The calculated spectrum described by Eq.({\ref{ur}}) at $Z=2$ and
$A/Z=5$.} \label{soho}
\end{figure}

Another spectrum was obtained in the program  "SOHO/GOLF".
Conversely, it is not characterized by high resolution, instead it
gives information about general character of the solar oscillation
spectrum (Fig.{\ref{soho}})).

The existence of this spectrum requires  to change the view at all
problems of solar oscillations. The theoretical explanation of this
spectrum must give answers at least to four questions :

1.Why does the whole spectrum consist from a large number of
equidistant spectral lines?

2.Why does the central frequency of this spectrum ${\cal F}$ is
approximately equal  to $\approx 3.23~ mHz$?

3. Why does  this spectrum splitting $f$ is approximately equal  to
$67.5~ \mu Hz$?

4. Why does the intensity of spectral lines decrease  from the
central line to the periphery?

The answers to these questions can be obtained if we take into
account  electric polarization of a solar core.

The description  of measured spectra by means of spherical analysis
does not make clear of the physical meaning of this procedure. The
reason of difficulties lies in attempt to consider the oscillations
of a Sun as a whole. At existing dividing of a star into core and
atmosphere, it is easy to understand that the core oscillation must
form a measured spectrum. The fundamental mode of this oscillation
must be determined by its spherical mode when the Sun radius
oscillates without changing of the spherical form of the core. It
gives a most low-lying mode with frequency:
\begin{equation}
\Omega_s\approx \frac{c_s}{\mathbb{R}_\star},
\end{equation}
where $c_s$ is sound velocity in the core.

It is not difficult to obtain the numerical estimation of this
frequency by order of magnitude. Supposing that the sound velocity
in dense matter is $10^7 cm/c$ and radius is close to
$\frac{1}{10}$ of external radius of a star, i.e. about $10^{10}cm$,
one can obtain as a result
\begin{equation}
F_s=\frac{\Omega_s}{2\pi}\approx 10^{-3}Hz
\end{equation}
It gives possibility to conclude that this estimation is in
agreement with measured frequencies. Let us consider this mechanism
in more detail.

\section[The sound speed in plasma]{The sound speed in hot plasma}

The pressure of high temperature plasma is a sum of the plasma
pressure (ideal gas pressure) and the pressure of black radiation:
\begin{equation}
P=n_e kT +\frac{\pi^2}{45\hbar^3 c^3}(kT)^4.\label{pr2}
\end{equation}
and its entropy is
\begin{equation}
S=\frac{1}{\frac{A}{Z}m_p}~ln~\frac{(kT)^{3/2}}{n_e}
+\frac{4\pi^2}{45\hbar^3 c^3n_e}(k T)^3,\label{s}
\end{equation}
The sound speed $c_s$ can be expressed by Jacobian \cite{LL}:
\begin{equation}
c_s^2=\frac{D(P,S)}{D(\rho,S)}=\frac{\biggl(\frac{D(P,S)}{D(n_e,T)}\biggr)}{\biggl(\frac{D(\rho,S)}{D(n_e,T)}\biggr)}
\end{equation}
or
\begin{equation}
c_s=\biggl\{\frac{5}{3}\frac{k{T}}{A/Z m_p}
\biggl[1+\frac{2\left(\frac{4\pi^2}{45\hbar^3
c^3}\right)^2(kT)^6}{5n_e[n_e+\frac{8\pi^2}{45\hbar^3
c^3}(kT)^3]}\biggr]\biggr\}^{1/2}
\end{equation}
For  $T={\mathbb{T}_\star}$ and $n_e=n_\star$ we have:
\begin{equation}
{\frac{4\pi^2(k{\mathbb{T}_\star})^3}{45\hbar^3
c^3n_\star}}=\approx0.58~.
\end{equation}
 Finally we obtain:
\begin{equation}
c_s=\biggl\{\frac{5}{3}\frac{\mathbb{T}_\star}{(A/Z)m_p}[1.085]\biggr)^{1/2}
\approx 5.63~10^7~\biggl(\frac{Z +1}{A/Z}\biggr)^{1/2}~
cm/s~.\label{cs}
\end{equation}
\section[The basic elastic oscillation]{The basic elastic oscillation of a spherical core}
Star cores consist of dense high temperature plasma which is a
compressible matter. The basic mode of elastic vibrations of a
spherical core is related with its radius oscillation. For the
description of this type of oscillation, the potential $\phi$ of
displacement velocities $v_r=\frac{\partial \psi}{\partial r}$ can
be introduced  and the motion equation can be reduced to the wave
equation expressed through $\phi$ \cite{LL}:
\begin{equation}
c_s^2\Delta\phi=\ddot\phi,
\end{equation}
and a spherical derivative for periodical in time oscillations
$(\sim e^{-i\Omega_s t})$
 is:
\begin{equation}
\Delta\phi=\frac{1}{r^2}\frac{\partial}{\partial
r}\biggl(r^2\frac{\partial \phi}{\partial
r}\biggr)=-\frac{\Omega_s^2}{c_s^2}\phi~.
\end{equation}
It has the finite solution for the full core volume including its
center
\begin{equation}
\phi=\frac{A}{r}sin \frac{\Omega_s r}{c_s},
\end{equation}
where $A$ is a constant. For  small oscillations, when displacements
on the surface $u_R$ are small $(u_R/R=v_R/ \Omega_s R\rightarrow
0)$ we obtain the equation:
\begin{equation}
tg \frac{\Omega_s {\mathbb{R}}}{c_s}=\frac{\Omega_s
{\mathbb{R}}}{c_s}
\end{equation}
which has the solution:
\begin{equation}
\frac{\Omega_s {\mathbb{R}}}{c_s}\approx4.49.
\end{equation}
Taking into account  Eq.({\ref{cs}})), the main frequency of the
core radial elastic oscillation is
\begin{equation}
\Omega_s =
4.49\biggl\{\frac{10.85}{(3/2)^7\pi}\biggl[\frac{Gm_p}{r_B^3}\biggr]\frac{A}{Z}
\biggl(Z +1\biggr)^3\biggr\}^{1/2}.\label{qb}
\end{equation}
It can be seen that this frequency depends on ${Z}$ and ${A}/{Z}$
only.

Some values of  frequencies of radial sound oscillations ${\cal
F}=\Omega_s/2\pi$  calculated from this equation for selected $A/Z$
at $Z=1$ and $Z=2$ are shown in third column of Table
({\ref{st-osc}}).

{\hspace{8cm} Table({\ref{st-osc}}).}

\begin{tabular}{||c|c|c||c|c||}\hline\hline
&&${\cal F},mHz$&&${\cal F},mHz$\\
Z&A/Z&(calculated&star&\\
& & on Eq.({\ref{qb}}))
&&measured\\
\hline 1&1&0.79&$\eta~Bootis$&0.85\\ \hline
&&& The Procion$(A\alpha~CMi)$&1.04\\
1&2&1.11& & \\
&&&$\beta~Hydri$&1.08\\ \hline 2&2&2.03&&\\\hline
2&3&2.48&$\alpha~Cen~A$&2.37\\ \hline
2&4&2.87&&\\
\hline 2&5&\bf3.24&The Sun&\bf3.23\\ \hline\hline
\end{tabular}\label{st-osc}

\bigskip

The measured frequencies of surface vibrations for some stars
\cite{CD} are shown in right part of this table. The data for $\nu
~Indus$ and $\xi~Hydrae$  also exist \cite{CD}, but characteristic
frequencies of these stars are below 0.3 mHz and they have some
other mechanism of excitation, probably. One can conclude from the
data of Table 1 that the core of the Sun is basically composed by
helium-10. It is not a confusing conclusion, because  pressure which
exists inside the solar core amounts to $10^{19} {dyne}/{sm^2}$ and
it is capable to induce the neutronization process in plasma and to
stabilize neutron-excess nuclei.

\section[The low frequency oscillation]{The low frequency oscillation of the density of a neutral
plasma}

Hot plasma has the density ${n_\star}$ at its equilibrium state. The
local deviations from this state induce processes of density
oscillation since plasma tends to return to its steady-state
density. If we consider small periodic oscillations of core radius
\begin{equation}
R={\mathbb{R}}+ u_R \cdot sin~ \omega_{n_\star}t,
\end{equation}
where a radial displacement of plasma particles is small
($u_R\ll{\mathbb{R}}$), the oscillation process of plasma density
can be described by the equation

\begin{equation}
\frac{d{\mathcal{E}}}{dR}=\mathbb{M}\ddot R~.
\end{equation}
Taking into account
\begin{equation}
\frac{d{\mathcal{E}}}{dR}=\frac{d\mathcal{E}_{plasma}}{dn_e}\frac{dn_e}{dR}
\end{equation}
and
\begin{equation}
\frac{3}{8}\pi^{3/2} \mathbb{N}_e \frac{e^3 a_0^{3/2}}
{(k\mathbb{T})^{1/2}}\frac{n_\star}{\mathbb{R}^2}=\mathbb{M}\omega_{n_\star}^2
\end{equation}
From this we obtain
\begin{equation}
\omega_{n_\star}^2=\frac{3}{\pi^{1/2}}
k{\mathbb{T}}\biggl(\frac{e^2}{a_B k{\mathbb{T}}}\biggr)^{3/2}
\frac{(1+Z)^3}{{\mathbb{R}}^2 A/Z m_p}
\end{equation}\label{qm2}
and finally
\begin{equation}
\omega_{n_\star}=\biggl\{\frac{2^8}{3^5}\frac{{\pi}^{1/2}}{{10}^{1/2}}
\alpha^{3/2}\biggl[\frac{Gm_p}{a_B^3}\biggr]\frac{A}{Z}\biggl[Z+1\biggr]^{4.5}\bigg\}^{1/2},\label{qm3}
\end{equation}
where $\alpha=\frac{e^2}{\hbar c}$ is the fine structure constant.
These low frequency oscillations of neutral plasma density are
similar to phonons in solid bodies. At that oscillations with
multiple frequencies $k\omega_{n_\star}$ can exist. Their power is
proportional to $1/{\kappa}$, as the occupancy these levels in
energy spectrum must be reversely proportional to their energy
$k\hbar\omega_{n_\star}$. As result, low frequency oscillations of
plasma density constitute set of vibrations
\begin{equation}
\sum_{\kappa=1} \frac{1}{\kappa}~sin(\kappa\omega_{n_\star}t)~.
\end{equation}

\section[The spectrum of solar oscillations]{The spectrum of solar core oscillations}
The set of the low frequency oscillations with $\omega_\eta$ can be
induced by sound oscillations with $\Omega_s$. At that,
displacements obtain the spectrum:
\begin{equation}
u_R\sim \sin~\Omega_s t\cdot\sum_{\kappa=0}
\frac{1}{\kappa}~\sin~\kappa\omega_{n_\star}t\cdot \sim
\xi\sin~\Omega_s t + \sum_{\kappa=1} \frac{1}{\kappa}~\sin~(\Omega_s
\pm \kappa \omega_{n_\star} )t,\label{ur}
\end{equation}
where $\xi$ is a coefficient $\approx 1$.

This spectrum is shown in Fig.({\ref{soho}}).

The central frequency of experimentally measured distribution of
solar oscillations  is approximately equal to (Fig.({\ref{bison}}))
\begin{equation}
{\cal F}_\odot\approx 3.23~ mHz\label{ffs}
\end{equation}
and the experimentally measured frequency splitting in this spectrum
is approximately equal to
\begin{equation}
{f}_\odot\approx 68~ \mu Hz.\label{ffg}
\end{equation}
 A good agreement of the calculated frequencies of basic modes of
oscillations (from Eq.({\ref{qb}}) and Eq.({\ref{qm}})) with
measurement can be obtained at $Z=2$ and $A/Z=5$:
\begin{equation}
{\cal F}_{_{_{Z=2;\frac{A}{Z}=5}}} =
\frac{\Omega_s}{2\pi}=3.24~mHz;~f_{_{_{Z=2;\frac{A}{Z}=5}}}=\frac{\omega_{n_\star}}{2\pi}=68.1~\mu
Hz.
\end{equation}

\chapter[The energy generation ]{The energy generation and the time of life of the Sun}
\label{Ch10}

Now it is commonly accepted to think that the energy generation in
stars is basically caused by thermonuclear reactions of
hydrogen-helium cycles. It seems to be valid for heavy stars
consisting of hydrogen and helium at ratio $A/Z=1\div 2$. In this
part of the star mass spectrum (Fig.{\ref{starM}}), sharp lines are
absent (beside one related to He-3, probably). This part of spectrum
is rather smeared. But this conception is in contradiction with the
fact of an existence of the lined spectrum mass of stars with
$A/Z>3$. Reaction of this type must go with a change of relation
$A/Z$ of a nuclear fuel and "smearing" of narrow peaks of the
spectrum of star masses during milliards of years. It seems  that it
is possible to make agreement of the measured lined spectrum of
these stars and thermonuclear mechanism of reaction, if we suppose
that a basic thermonuclear reaction is
\begin{equation}
{_Z^AX}+{_Z^AX}={_{2Z}^{2A}X} + \gamma
\end{equation}
i.e., for example, two nuclei of tritium join up into a helium-6
nucleus:
\begin{equation}
{_1^3{H}}+{_1^3H}={_2^6He}+ \gamma
\end{equation}
For Sun, a process of nuclear fusion of two nuclei of hydrogen
$^5_1H$ joining into helium $^{10}_2He$ must be prevailing:
\begin{equation}
{_1^5{H}}+{_1^5H}={_2^{10}He}+ \gamma
\end{equation}
The mass of nucleus hydrogen $^5_1H$ is equal approximately to
$5.03954$ a.m.u., the mass of helium $^{10}_2He$ is equal
approximately to $10.0524$ à.å.ì. Thus, the energy about  $4\cdot
10^{-5}$ erg will be emitted in a single reaction. The full number
of nuclei in the star core:
\begin{equation}
\mathbb{N}_\star = \frac{\mathbb{M}_\star}{\frac{A}{Z}m_p}\approx
10^{56}.
\end{equation}

Now the Sun radiates from its surface  $3.86\cdot 10^{33}$ erg/s.
Approximately $10^{38}$ reactions during one second can provide for
this energy. The question emerges: how much hydrogen is there stored
into the Sun now?

The answer can be obtained from the analysis of the solar
oscillation frequencies. These measurements give two frequencies.
Their theoretical dependencies on chemical parameters $A/Z$ and $Z$
are known from Eqs.({\ref{qb}}) and ({\ref{qm2}}). Using these
equations, it is possible to express these two chemical parameters:
\begin{equation}
\langle A/Z\rangle =2.33\cdot 10^{-4}\frac{\pi^6 \alpha^3 r_B^3}{G
m_p} \biggl(\frac{{\cal F}^3} {f_\eta^2}\biggr)^{2}\label{a/z},
\end{equation}
\begin{equation}
\langle Z\rangle = \frac{11.14}{\pi \alpha}\biggl(\frac{f}{{\cal
F}}\biggr)^{4/3}-1\label{z}.
\end{equation}
The substitution of numerical values gives
\begin{equation}
\langle A/Z\rangle_\odot = 5.04~.\label{a/zz}
\end{equation}
and
\begin{equation}
\langle Z\rangle_\odot = 1.85\label{zz}
\end{equation}
The last equation speaks that now the Sun must consist approximately
from $85\%$ of $_2^{10}He$ and $15\%$ of $_1^5H$.

Experts think that approximately for the last 5 milliard years the
Sun was shining more or less monotone  without catastrophic jumps of
radiation.

It is not difficult to estimate, that stockpiles  of nuclear fuel
$_1^5H$ at a present speed  burning must be enough to the Sun for
several milliard years ahead.

This consideration extremely radically "solves" the problem of solar
neutrinos. Reactions considered above  yield no neutrinos at all. In
this connection, the question emerges: is it possible to agree the
considered mechanism based on the form of stellar mass spectrum with
the measured flux of neutrinos?

\chapter[Other stars]
{Other stars, their classification and some cosmology} \label{Ch11}

The Schwarzsprung-Rassel diagram is now a generally accepted base
for star classification. It seems that a classification based on the
EOS of substance may be more justified from physical point of view.
It can be  emphasized by possibility to determine the number of
classes of celestial bodies.

The matter can totally have seven states

The atomic substance at low temperature exists as  condensed matter
(solid or liquid). At high temperature it transforms into gas phase.

The electron -nuclear plasma can exist in four states. It can be
relativistic or non-relativistic. The electron gas of
non-relativistic plasma can be degenerate (cold) or non-degenerate
(hot). If electron gas of plasma is relativistic, its nuclear
subsystem may be cold at low temperature. At very high temperature,
the energy of nuclear subsystem can exceed the energy of
degeneration of relativistic electrons.

In addition to that a substance can exist as  neutron matter with
nuclear density in degenerate state.

At present,  assumptions about existence of matter at different
states, other than the above-named,  seem unfounded. Thus, seven
possible states of matter show a possibility of classification  of
celestial bodies in accordance with this dividing.

\section{The atomic substance}
\subsection{Small bodies} Small celestial bodies - asteroids and
satellites of planets - are usually considered as bodies consisting
from atomic matter.

\subsection{Giants} The transformation of atomic matter into plasma can
be induced by action of high pressure, high temperature or both
these factors. If these factors inside a body are not high enough,
atomic substance can transform into gas state. The characteristic
property of this celestial body is absence of electric polarization
inside it. If temperature of a body is below ionization temperature
of atomic substance but higher than its evaporation temperature, the
equilibrium equation comes to
\begin{equation}
-\frac{dP}{dr}=\frac{G\gamma}{r^2}M_r\approx\frac{P}{R}\approx
\frac{\gamma}{m_p} \frac{kT}{R}.
\end{equation}
Thus, the radius of the body
\begin{equation}
R \approx \frac{GM m_p}{kT}.
\end{equation}
If its mass $M\approx 10^{33}~g$ and temperature at its center
$T\approx 10^5~K$, its radius is  $R\approx 10^2~R_{\odot}$. These
proporties are characteristic for giants, where pressure at center
is about $P\approx 10^{10}~din/cm^2$ and it is not enough for
substance ionization.

\section{Plasmas}
\subsection[Stars]{The non-relativistic non-degenerate plasma. Stars.}
Characteristic properties of hot stars consisting of
non-relativistic non-degenerate plasma was considered above in
detail. Its EOS is ideal gas equation.

\subsection[Planet]{Non-relativistic degenerate plasma. Planets.}
At cores of large planets, pressures are large enough to transform
their substance into plasma. As temperatures are not very high here,
it can be supposed, that  this plasma can  degenerate:
\begin{equation}
T<<T_F
\end{equation}
The pressure which is induced by gravitation must be balanced by
pressure of non-relativistic degenerate electron gas
\begin{equation}
\frac{G\mathbb{M}^2}{6\mathbb{R}\mathbb{V}}\approx\frac{(3\pi^2)^{2/3}}{5}
\frac{\hbar^2}{m_e} \biggl(\frac{\gamma}{m_p A/Z}\biggr)^{5/3}
\end{equation}
It opens  a way to estimate the mass of this body:
\begin{equation}
\mathbb{M}\approx\mathbb{M}_{Ch}
\biggl(\frac{\hbar}{mc}\biggr)^{3/2}
\biggl(\frac{\gamma}{m_p}\biggr)^{1/2}
\frac{6^{3/2}9\pi}{4(A/Z)^{5/2}}
\end{equation}

At density about  $\gamma\approx 1~g/cm^3 $, which is characteristic
for large planets, we obtain their masses
\begin{equation}
\mathbb{M}\approx 10^{-3} \frac{\mathbb{M}_{Ch}}{(A/Z)^{5/2}}
\approx\frac{4\cdot 10^{30}}{(A/Z)^{5/2}} ~ g
\end{equation}
Thus, if we suppose that large planets consist of hydrogen (A/Z=1),
their masses must not be over  $4\cdot 10^{30}g$. It is in agreement
with the Jupiter's mass, the biggest planet of the Sun system.

\subsection{The cold relativistic substance}
An energy of a relativistic fermi-system with volume $V$ can be
obtained from the general expression \cite{LL}:
\begin{equation}
\mathcal{E}=\frac{Vc}{\pi^2 \hbar^3}\int_0^{p_F}p^2\sqrt{m^2
c^2+p^2}dp.
\end{equation}
At integrating of this expression after subtracting of self-energy
of particles, we obtain the kinetic energy of $N$ relativistic
particles:
\begin{equation}
\mathcal{E}_{kin}=\frac{3}{8}Nmc^2
\left[\frac{x(2x^2+1)\sqrt{x^2+1}-Arcsinh(x)-\frac{8}{3}x^3}{x^3}\right]
\label{e-rs}
\end{equation}
where $x=\frac{p_F}{mc}$ and density of substance:
\begin{equation}
n=\frac{p_F^3}{3\pi^2\hbar^3}=\frac{x^3}{3\pi^2}
\left(\frac{mc}{\hbar}\right)^3.\label{nn-rs}
\end{equation}
The pressure of this system is
\begin{equation}
P=-\left(\frac{d\mathcal{E}_{kin}}{dV}\right)_{S=0}=\frac{mc^2}{8\pi^2}
\left(\frac{mc}{\hbar}\right)^3
\left[x\left(\frac{2}{3}x^2-1\right)\sqrt{x^2+1}+Arcsinh(x)\right]
\end{equation}
For simplicity we can suppose that main part of mass of relativistic
star is concentrated into its core and it is uniformly distributed
here. In this case we can write the pressure balance equation:
\begin{equation}
\frac{GM^2}{2R}=3PV.
\end{equation}
It permits to estimate the full number of particles into the
considered system:
\begin{equation}
N=\left(\frac{\hbar
c}{Gm_p^2}\right)^{3/2}\biggl(\frac{3}{4\pi}\biggr)^{2}\frac{1}{\pi^{3/2}}
\left[\frac{x\left(\frac{2}{3}x^2-1\right)\sqrt{x^2+1}+Arcsinh(x)}{x^4}\right]^{3/2}.
\label{m-puls}
\end{equation}
Accordingly to the virial theorem:
\begin{equation}
\mathcal{E}_{total}=-\mathcal{E}_{kin}
\end{equation}
and ({\ref{e-rs}}), the full energy of a relativistic star
\begin{eqnarray}
\mathcal{E}_{total}\sim
-\frac{1}{x^9}
\left[x(2x^2+1)\sqrt{x^2+1}-Arcsinh(x)-\frac{8}{3}x^3\right]\cdot
\nonumber \\
\cdot\left[x\left(\frac{2}{3}x^2-1\right)\sqrt{x^2+1}+Arcsinh(x)\right]^{3/2}.\label{e-rt}
\end{eqnarray}
Because this expression has a minimum at $x\approx 1.35$, a star
consisting of relativistic substance must have the equilibrium
particle density:
\begin{equation}
n_{\star}\approx 8.3\cdot10^{-2}\left(\frac{mc}{\hbar}\right)^3.
\label{n-rs}
\end{equation}

\subsubsection[Dwarfs]{The relativistic degenerate plasma. Dwarfs.}

The degenerate relativistic plasma has a cold relativistic electron
subsystem. Its nuclear subsystem can be non-relativistic. The star
consisting of this plasma accordingly to Eq.({\ref{n-rs}}) at
$m=m_e$, must have an energy minimum at electron density
$n_\star\approx 1. 4\cdot 10^{30}$  $particle/cm^3$ at a star radius
\begin {equation}
R\approx {10^{-2}R_{\odot}}
\end {equation}
It is not difficult to see that these densities and radii are
characteristic for dwarfs.

\subsubsection[Pulsars]{The neutron matter. Pulsars.} Dwarfs may be
considered as stars where a process of neutronization is just
beginning. At a nuclear density, plasma turns into neutron matter.
\footnote{At nuclear density neutrons and protons are
indistinguishable inside pulsars as inside a huge nucleus. It
permits to suppose a possibility of gravity induced electric
polarization in this matter. }

In accordance with Eq.({\ref{n-rs}}) at $m=m_n$, a star consisting
of neutron matter must have the minimal energy at neutron density
$n_\star\approx 9\cdot 10^{39}$ $particle/cm^3$. As result the
masses of neutron stars must be  approximately equal to $M_{Ch}$.
The measured mass distribution of pulsar composing binary stars
\cite{Thor} is shown on Fig.\ref{pulsar}. It can be considered as a
confirmation of the last conclusion.

\begin{figure}
\begin{center}
\centering\includegraphics[height=10cm]{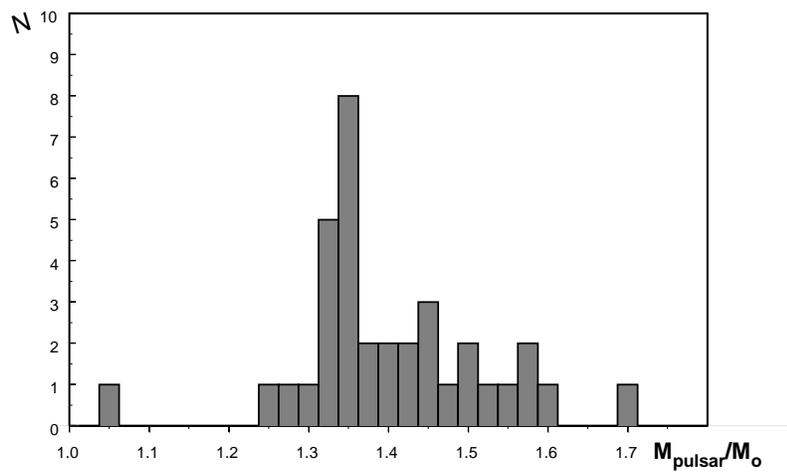} \caption{The
mass distribution of pulsars from binary systems \cite{Thor}.On
abscissa the logarithm of pulsar mass in solar mass is shown.}
\label{pulsar}
\end{center}
\end{figure}

\subsection[Quasars?]{The hot relativistic plasma.Quasars?}
Plasma is hot if its temperature  is higher than degeneration
temperature of its electron gas. The ratio of plasma temperature in
the core of a star to the temperature of degradation of its electron
gas for case of non-relativistic hot star plasma is
(Eq.({\ref{alf}}))
\begin{equation}
\frac{\mathbb{T}_\star}{T_F(n_\star)} \approx 40
\end{equation}
it can be supposed that the same ratio must be characteristic for
the case of a relativistic hot star. At this temperature, the
radiation pressure plays a main role and accordingly the equation of
the pressure balance takes the form:

\begin{equation}
\frac{GM^2}{6RV}\approx\frac{\pi^2}{45}\frac{(k\mathbb{T}_\star)^4}{(\hbar
c)^3}\approx \biggl(\frac{\mathbb{T}_\star}{T_F}\biggr)^3 kTn
\end{equation}

\begin{figure}
\begin{center}
\centering\includegraphics[scale=0.5]{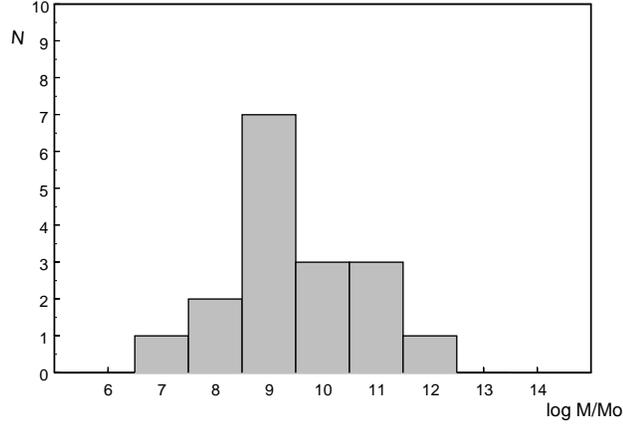}
\caption{The mass distribution of galaxies. \cite{Al}. On the
abscissa, the logarithm of the galaxy mass over the Sun mass is
shown.} \label{galaxy}
\end{center}
\end{figure}

This makes it possible to estimate the mass of a hot relativistic
star
\begin{equation}
M\approx \biggl(\frac{\mathbb{}T_\star}{T_F}\biggr)^6
\biggl(\frac{\hbar c}{G m_p^2} \biggr)^{3/2} m_p \approx 10^9
M_{\odot}
\end{equation}
According to the existing knowledge, among compact celestial objects
only quasars have masses of this level. Apparently it is an
agreed-upon opinion that quasars represent some relatively short
stage of evolution of galaxies. If we adhere to this hypothesis, the
lack of information about quasar mass distribution can be replaced
by the distribution of masses of galaxies
\cite{Al}(Fig.{\ref{galaxy}}). It can be seen, that this
distribution is in a qualitative agreement with supposition that
quasars are composed from the relativistic hot plasma.

\section{Some cosmology}
Thus,  it seems possible under some assumptions to find
characteristic parameters of different classes of stars, if to
proceed from EOS of atomic, plasma and neutron substances. Seven EOS
can be compared with seven classes of celestial bodies. As any other
EOS are unknown, it gives a reason to think that all classes of
celestial bodies are discovered. With the exception of ,probably,
one body that could have existed in  the past. The neutron matter at
density of order $10^{40} particle/cm^3$ (in accordance with
Eq.({\ref{n-rs}})) is not ultra-relativistic matter and its energy
and pressure must depend on temperature.\footnote{The
ultra-relativistic matter with $p_F\gg mc$ is possessed  by limiting
pressure which is not depending on temperature.} It seems, there is
no thermodynamical prohibition to imagine this matter so hot when
the black radiation pressure will dominate inside it. An estimation
shows that it can be possible if mass of this body is near to
$10^{50}g$ or even to $10^{55}g$. As it is accepted to think that
full mass of the Universe is about $10^{53}g$, it can be assumed
that on an early stage of development of Universe, there was some
body with a mass about  $10^{53}g$ composed by the neutron matter at
nuclear density with the radiation at temperature above $10^{12}K$.
After some time, with temperature decreased it has lost its
stability and decayed into quasars with mass up to $10^{12}M_{Ch}$,
consisting of a relativistic plasma with hot nuclear component at
$T>10^{10}K$. After next cooling at loosing of stability they was
decaying on galaxies of hot stars with mass about $M\approx M_{Ch}$
and core temperature about $T\approx 10^{7}K$, composed by
non-relativistic hot plasma. A next cooling must leads hot stars to
decaying on dwarfs, pulsars, planets or may be on small bodies. The
substances of these bodies (in their cores) consists of degenerate
plasma (degenerate electron subsystem and cold nuclear subsystem) or
cold neutron matter, it makes them stable in expanding and cooling
Universe.\footnote{The temperature of plasma inside these bodies can
be really quite high as electron gas into dwarfs, for example, will
be degenerate even at temperature $T\approx 10^{9}K$.}

\chapter
{The conclusion} \label{Ch12}

Evidently, the  main conclusion from the above consideration
consists in statement of the fact that now there are quite enough
measuring data to place the theoretical astrophysics on a reliable
foundation. All above measuring data are known  for a relatively
long time. The traditional system of view based on the Euler
equation in the form ({\ref{Eu}}) could not give a possibility to
explain  and even to consider. with due proper attention, to  these
data. Taking into account  the gravity induced electric polarization
of plasma and  a change starting postulate gives a possibility to
obtain results for  explanation of
 measuring data considered above.

\hspace{1cm}

Basically these results are the following.

\hspace{1cm}

Using the standard method of plasma description leads to the
conclusion that at conditions characteristic for the central stellar
region, the plasma has the minimum energy  at constant density
$n_\star$ (Eq.(\ref{eta1})) and at the constant temperature
$\mathbb{T}_\star$ (Eq.(\ref{tcore})).

\hspace{1cm}

This plasma forms the core of a star, where the pressure is constant
and gravity action is balanced by the force of the gravity induced
by the electric polarization. The virial theorem gives a possibility
to calculate the stellar core mass $\mathbb{M}_\star$
(Eq.(\ref{Mcore})) and its radius $\mathbb{R}_\star$
({\ref{Rcore}}). At that the stellar core volume is approximately
equal to  1/1000 part of full volume of a star.

\hspace{1cm}

The remaining mass of a star located over the core has a density
approximately thousand times smaller and it is convenient to name it
a star atmosphere. At using thermodynamical arguments, it is
possible to obtain the radial dependence of plasma density inside
the atmosphere $n_a\approx r^{-6}$ (Eq.({\ref{an-r}})) and the
radial dependence of its temperature $\mathbb{T}_a\approx r^{-4}$
(Eq.({\ref{tr}})).

\hspace{1cm}

It gives a possibility to conclude that  the mass of the stellar
atmosphere  $\mathbb{M}_a$ (Eq.({\ref{ma}})) is almost exactly equal
to the stellar core mass. Thus, the full stellar mass can be
calculated. It depends on the ratio of the mass and the charge of
nuclei composing the plasma. This claim is in a good agreement with
the measuring data of the mass distribution of both -  binary stars
and close binary stars  (Fig.
({\ref{starM}})-({\ref{starM2}}))\footnote{The measurement of
parameters of these stars has a satisfactory  accuracy only.}. At
that it is important that the upper limit of masses of both - binary
stars and close binary stars - is in accordance   with the
calculated value of the mass of the hydrogen star (Eq.({\ref{M}})).
The obtained formula explains the origin of sharp peaks of stellar
mass distribution - they evidence that the substance of these stars
have a certain value of the ratio  $A/Z$. In particular the solar
plasma according to  (Eq.({\ref{M}})) consists of nuclei with
$A/Z=5$.

\hspace{1cm}

 Knowing temperature and substance density on the core and knowning their radial dependencies, it is possible to estimate the surface temperature  $\mathbb{T}_0$ (\ref{T0})
and the radius of a star $\mathbb{R}_0$ ({\ref{R0}}). It turns out
that these measured parameters must be related to the star mass with
the  ratio $\mathbb{T}_0 \mathbb{R}_0\sim \mathbb{M}^{5/4}$
({\ref{5/4}}). It is in a good agreement with measuring data
(Fig.({\ref{RT-M}})).

\hspace{1cm}

Using another thermodynamical relation - the Poisson's adiabat -
gives a way to determine the relation between radii of stars and
their masses $\mathbb{R}_0^3\sim \mathbb{M}^2$ (Eq.({\ref{rm23}})),
and between their surface temperatures and masses  $\mathbb{T}_0\sim
\mathbb{M}^{5/7}$ (Eq.({\ref{tm}})). It gives the quantitative
explanation of the mass-luminosity dependence (Fig.({\ref{LM}})).

\hspace{1cm}

According to another familiar Blackett's dependence, the
giromagnetic ratios of celestial bodies are approximately equal to
$\sqrt{G}/c$. It has a simple explanation too. When there is the
gravity induced electric polarization of a substance of a celestial
body, its rotation must induce a magnetic field
(Fig.({\ref{black}})). It is important that all relatively large
celestial bodies - planets, stars, pulsars - obey the Blackett's
dependence. It confirms a consideration that the gravity induced
electric polarization must be characterizing for all kind of plasma.
The calculation of magnetic fields of hot stars shows that they must
be proportional to rotation velocity of stars  ({\ref{Ht}}).
Magnetic fields of Ap-stars are measured, and they can be compared
with periods of changing of luminosity of these stars. It is
possible that this mechanism is characteristic for stars with rapid
rotation (Fig.({\ref{H-W}})), but obviously there are other
unaccounted factors.

\hspace{1cm}

Taking into account the gravity induced electric polarization and
coming from the Clairault's theory, we can describe the periastron
rotation of binary stars as effect descended from non-spherical
forms of star cores. It gives the quantitative explanation of this
effect, which is in a good agreement with measuring data
(Fig.({\ref{periastr}})).

\hspace{1cm}

The solar oscillations can be considered as elastic vibrations of
the solar core. It permits to obtain two basic frequencies of this
oscillation: the basic frequency of sound radial oscillation of the
core and the frequency of splitting depending on oscillations of
substance density near its equilibrium value. It yeils a good
agreement with the measuring data and demonstrates that the Sun
consists generally of helium-10 (Fig.({\ref{soho}})).

\hspace{1cm}

A calculation can be carried out in reverse direction. Two measured
frequencies of solar oscillations allow  to calculate the chemical
composition of solar substance. It must be composed by 85\% of
helium-10 and by 15\% of hydrogen-5. It shows that the process of
quiet burning of the Sum will be continue for a few milliards years
ahead.

\hspace{1cm}

The plasma can exists in four possible states. The  non-relativistic
electron gas of plasma can be degenerate and non-degenerate. Plasma
with relativistic electron gas can have a cold and a hot nuclear
subsystem.  Together with the atomic substance and neutron
substance, it gives seven possible states. It suggests a  way of a
possible classification of celestial bodies. The advantage of this
method of classification is in the possibility to estimate
theoretically main parameters characterizing the celestial bodies of
each class. And these predicted parameters are in agreement with
astronomical observations. It can be supposed hypothetically  that
cosmologic transitions between these classes go in direction of
their temperature being lowered. But these suppositions have no
formal base at all.

\hspace{1cm}

Discussing  formulas obtained, which  describe star properties, one
can note a important moment: these considerations  permit to look at
this problem from different points of view. On the one hand, the
series of conclusions follows from existence of spectrum of star
mass (Fig.({\ref{starM}})) and from known chemical composition
dependence. On the another hand, the calculation of natural
frequencies of the solar core gives a different approach to a
problem of chemical composition determination. It is important that
for the Sun, both these approaches lead to the same conclusion
independently and unambiguously. It gives a confidence in
reliability of obtained results.

\hspace{1cm}

In fact, the calculation of magnetic fields of hot stars only does
not give a satisfactory explanation of existing measuring data. At
that  it is unlikely that the reason of discrepancy is in the use of
dubious suggestions - the dipole magnetic field of the Sun is not
explained too. But it is important, that the all remaining measuring
data  (all known of today) confirm both - the new postulate and
formulas based on it. At that, the main stellar parameters - masses,
radii and temperatures - are expressed through combinations of world
constants and at that they show a good accordance with observation
data. It is important  that quite a satisfactory quantitative
agreement of obtained results and measuring data can be achieved by
simple and clear physical methods without use of any fitting
parameter. It gives a special charm and attractiveness to  star
physics.

\clearpage



\end{document}